\documentclass{article} 

\usepackage{iclr2027_conference,times}

\usepackage[utf8]{inputenc} 
\usepackage[T1]{fontenc}    
\usepackage{hyperref}       
\usepackage{url}            
\usepackage{booktabs}       
\usepackage{amsfonts}       
\usepackage{nicefrac}       
\usepackage{microtype}      
\usepackage{xcolor}         

\usepackage{latexsym}
\usepackage{enumitem}

\usepackage{graphicx}
\usepackage{amsmath}
\usepackage{amssymb}
\usepackage{comment}

\usepackage{caption}
\usepackage{subcaption}

\usepackage{multicol}
\usepackage{multirow}
\usepackage{array, tabularx, ragged2e}

\usepackage{pgfplots}
\usepgfplotslibrary{groupplots}
\pgfplotsset{compat=1.18}

\usepackage{ctable}
\usepackage{wrapfig}
\usepackage{tikz}
\usetikzlibrary{arrows.meta,positioning,arrows,trees,shapes,fit,shadows,patterns}
\usepackage{iclr2027_arxiv}

\usepackage{ifthen}
\usepackage{supertabular}
\usepackage[normalem]{ulem}
\usepackage{hhline}
\usepackage{lettrine}

\usepackage{algorithm}
\usepackage[noend]{algpseudocode}

\makeatletter
\def\theHALG@line{\thealgorithm.\arabic{ALG@line}}
\makeatother

\usepackage{xspace}
\usepackage[export]{adjustbox}
\usepackage[capitalize,noabbrev]{cleveref}
\usepackage{pifont}
\usepackage{float}

\crefformat{section}{\S#2#1#3} 
\crefformat{subsection}{\S#2#1#3}
\crefformat{subsubsection}{\S#2#1#3}
\crefname{algorithm}{Alg.}{Algs.}

\usepackage{amsmath,amsfonts,bm}

\def\eqref#1{equation~\ref{#1}}

\def\1{\bm{1}}

\makeatletter
\newcommand{\xRightarrow}[2][]{\ext@arrow 0359\Rightarrowfill@{#1}{#2}}
\makeatother

\DeclareMathAlphabet{\mathsfit}{\encodingdefault}{\sfdefault}{m}{sl}
\SetMathAlphabet{\mathsfit}{bold}{\encodingdefault}{\sfdefault}{bx}{n}

\newcommand{\Ni}{({\em i})~}
\newcommand{\Nii}{({\em ii})~}
\newcommand{\Niii}{({\em iii})~}
\newcommand{\Niv}{({\em iv})~}

\newcommand{\gib}{\,\mathrm{GiB}}
\newcommand{\mib}{\,\mathrm{MiB}}

\newcommand{\ceff}{c_{\mathrm{eff}}}

\newcommand{\mop}{{MoP}\xspace}
\newcommand{\lleptwo}{{PipelinedLLEP}\xspace}
\newcommand{\ringtp}{{Ring-DTP}\xspace}
\newcommand{\selectoffload}{{SCO}\xspace}
\newcommand{\streamadam}{{OffloadStreamAdamW}\xspace}

\newcommand{\ringtplong}{{ring data-tensor-parallel vocabulary projection}\xspace}
\newcommand{\selectoffloadlong}{{selective checkpoint offload}\xspace}
\newcommand{\streamadamlong}{{streamed offloaded AdamW}\xspace}

\usepackage{mathtools}

\definecolor{routeblue}{HTML}{2864B7}
\definecolor{offloadorange}{HTML}{D97706}
\definecolor{baselinegray}{HTML}{87909B}
\definecolor{successgreen}{HTML}{238B57}
\definecolor{darkblue}{rgb}{0, 0, 0.5}

\definecolor{vocabpurple}{HTML}{7A4BA8}
\definecolor{warnred}{HTML}{A33B3B}

\tikzset{
  fglbl/.style={font=\fontsize{6}{7}\selectfont},
  fgtitle/.style={font=\scriptsize\bfseries, text=black!85},
  fgaxis/.style={draw=black!45, line width=0.4pt},
  fgbound/.style={draw=successgreen, line width=0.7pt,
                  dash pattern=on 2pt off 1.4pt},
  fgarrow/.style={-{Stealth[length=1.3mm]}, draw=black!55, line width=0.35pt},
}

\hypersetup{
    colorlinks=true,
    linkcolor=darkblue,
    citecolor=darkblue,
    urlcolor=darkblue,
    pdftitle={Flattening Every Memory Peak in Long-Context
              Mixture-of-Experts Training},
    pdfauthor={Shrey Pandit, Xuan-Phi Nguyen, Yiran Zhao, Shafiq Joty}
}

\setlist{leftmargin=*,nosep}

\newcolumntype{L}[1]{>{\raggedright\arraybackslash}p{#1}}
\newcolumntype{R}[1]{>{\raggedleft\arraybackslash}p{#1}}

\title{Flattening Every Memory Peak in Long-Context Mixture-of-Experts Training}

\author{%
    Xuan-Phi Nguyen$^{\dagger}$\thanks{Corresponding authors: \href{mailto:xnguyen@salesforce.com}{\{xnguyen,sjoty\}@salesforce.com}} \qquad Shrey Pandit$^{\dagger}$ \qquad Yiran Zhao$^{\dagger}$
    \\
    \qquad Shafiq Joty$^{*}$ \\
    Salesforce AI Research
}

\seticlrarxivauthors{%
  Shrey Pandit$^{\dagger}$, Xuan-Phi Nguyen$^{\dagger,*}$, 
  Yiran Zhao, Shafiq Joty$^{*}$%
}
\seticlrarxivaffiliation{Salesforce AI Research}
\seticlrarxivauthornote{%
  $^{\dagger}$Equal contribution.\quad
  $^{*}$Corresponding authors:
  \href{mailto:xnguyen@salesforce.com}{\{xnguyen,sjoty\}@salesforce.com}%
}

\iclrarxivcopy

\begin{document}

\maketitle

\begin{abstract}
Training a Mixture-of-Experts (MoE) model at long context or large batch size
fails as soon as any one component's peak allocation exceeds device memory, so
the target is every peak at once, not the average footprint.  Four are left
unbounded by the parallelism plans in common use, and each grows differently:
expert dispatch with the routing matrix, the vocabulary projection with tokens
times vocabulary, gradient checkpoint boundaries with depth times sequence
length, and optimizer state with parameter count.  Which one runs out first
changes with the model, the context length, and the device count, so lowering
the largest only exposes the next.  We bound all four with schedules whose GPU
working set is fixed at launch: PipelinedLLEP extends least-loaded expert parallelism
with a cap on the tokens each source contributes to a dispatch chunk, Ring-DTP
circulates activations or weight shards around a ring at the vocabulary
projection and folds each block of logits into an online log-sum-exp,
Selective checkpoint offload (SCO) keeps the one long-lived tensor of each
checkpoint boundary in CPU memory, and OffloadStreamAdamW turns the serial CPU
Adam update of optimizer offload into a bucket pipeline.  All four change only
the order and granularity of computation and data movement, so the loss and
gradients stay exact.  In matched component tests, they cut the MoE dispatch peak by
up to $59.3\%$ without losing throughput, the vocabulary projection peak by
$86.6\%$, and the offloaded optimizer step by $2.05\times$ faster.  Composed on MoE
models from 120B to 667B parameters, they train at 1M context length,
$8$--$32\times$ the reach of a tuned FSDP2 baseline, and up to $10.4\times$ its
throughput.
\end{abstract}

\section{Introduction}\label{sec:main:intro}

Mixture-of-Experts (MoE) models are often trained at long context or large batch size at large scale of compute. However, when adding GPUs is rarely the available remedy, memory is traded for time.  Every standard trade charges time: gradient checkpointing recomputes \citep{chen2016sublinear,korthikanti2023reducing}, state sharding communicates \citep{rajbhandari2020zero,zhao2023fsdp}, and optimizer offload serializes the update on the host \citep{ren2021zerooffload}. 
A training step dies when any component's peak memory occupancy exceeds the device's capacity. So a plan that reduces three bottlenecks and leaves the fourth free to grow with the workload buys nothing at the point where the workload grows. Reaching longer context, or larger batches, or large model sizes means holding every component under the device ceiling at the same time.

\paragraph{Which peak runs out first depends on the configuration.}
Peak memory is a maximum over \emph{live sets}, the groups of tensors that must be resident simultaneously. In this work, we explore four components that are left unbounded by the parallelism plans in common use: expert dispatch grows with the routing matrix the router draws \citep{gshard_lepikhin2020,rajbhandari2022deepspeedmoe}, the vocabulary projection with the token-vocabulary product \citep{wijmans2025cce}, retained gradient checkpointing boundaries with depth times sequence length \citep{chen2016sublinear}, and AdamW state with parameter count \citep{loshchilov2019decoupled}. Because these grow at different rates, the largest of the four depends specific configuration: logits dominate at large vocabulary and long context, MoE dispatch at high routing imbalance, checkpointing boundaries at high depth, and optimizer state at large parameter count on few devices. 

A usable stack needs a bound on every term rather than a large saving on one, and each bound has to be separately enableable so a training run pays only for the peaks it has, which is why the four operators below are developed together. All four change only the order and granularity of computation and data movement, leaving the model, the parameterization, the optimizer, the precision, and the loss untouched: no low-rank adapters, no quantized state \citep{dettmers2022eightbit}, and no approximate routing or attention. Memory and throughput figures are therefore directly comparable with standard full-parameter BF16 training.

\paragraph{Four bounded-streaming operators.}
Each operator takes one established component, identifies what that component leaves unbounded, and replaces its schedule.

\Ni Expert parallelism shards experts across ranks (GPUs), so a rank must hold every token routed to the experts it owns, and the router decides how many that is \citep{gshard_lepikhin2020,rajbhandari2022deepspeedmoe}.  Least-loaded expert parallelism (LLEP) removes the time cost of routing skew by moving routed work to idle ranks without changing any token's expert choice \citep{nguyen2026llep}, but a rank still allocates for its whole routed batch at once. Our proposed \lleptwo delivers the same routes in chunks and caps the tokens any source may place in a chunk, so a receiver's buffers follow that cap and routing skew lengthens the schedule instead of enlarging the buffers.

\Nii The vocabulary projection materializes a tokens-by-vocabulary logit tensor, which dominates memory once both factors are large.  Fused cross-entropy kernels avoid it on a single device by streaming the log-sum-exp over vocabulary blocks \citep{wijmans2025cce,hsu2024liger}, and Megatron shards the weight but requires every rank to hold the same batch \citep{shoeybi2019megatron}, reducing effective batch size.  Our \ringtp (\ringtplong) lets each rank keep its own distinct batch and circulates either the activations or the weight shards around a ring until every batch has met every shard, the exact loss and gradients are computed yet the large tensor is never formed.

\Niii Gradient checkpointing buys memory with recomputation, yet one tensor per checkpointed layer, its input, still lives from the forward pass until that layer is recomputed \citep{chen2016sublinear}. Our \selectoffload (\selectoffloadlong) offloads a host-budget subset of those inputs into host memory, and prefetches them back one layer before the layer is recomputed.

\Niv Optimizer offload stores AdamW state on CPU RAM, but a CPU update operation is slow, during which the GPU is idle and, having released its activations, nearly empty \citep{ren2021zerooffload}. Instead, our \streamadam (\streamadamlong) streams the parameters from host RAM to GPUs and compute the weight update there and write back. It leverages communication-computpation overlap to maximize the efficiency.

\paragraph{Individual improvements.}
Each operator is measured against the baseline that solves the same problem, on identical input (\cref{sec:main:method}).
\begin{itemize}
  \item \lleptwo lowers the MoE dispatch peak by up to $59.3\%$ against LLEP and runs with no noticeable slowdown against it.
  \item \ringtp removes $86.6\%$ of the vocabulary projection peak for under $5\%$ more time.
  \item \selectoffload lowers the device peak monotonically in the host budget it is given, moves throughput by under $2\%$, and raises the largest batch that completes by $17.7\%$.
  \item \streamadam runs the offloaded optimizer step $2.05\times$ faster than the CPU AdamW it replaces.
\end{itemize}

We compose the four operators inside the Mixture-of-Parallelisms (\mop) rank layout \citep{nguyen2026mop}.
End to end on MoE models of 120B, 241B, and 667B parameters, the composed stack trains at one-million context length, $8$--$32\times$ the reach of a tuned FSDP2 baseline, at up to $10.4\times$ its throughput and up to $12\times$ its largest global batch (\cref{sec:main:experiments}).

\section{Four Scaling Axes}\label{sec:main:framework}

Let $\Theta$ be the number of trainable parameters, $W$ the number of devices,
$N$ the tokens resident per rank, $H$ the hidden width, $V$ the vocabulary
size, and $b$ the bytes per activation scalar.  AdamW
\citep{loshchilov2019decoupled} ordinarily requires $16\Theta$ bytes for
working weights, gradients, master weights, and moments.  ZeRO-3 shards that
persistent state, sequence parallelism reduces attention activations, and
expert parallelism partitions expert weights
\citep{rajbhandari2020zero,zhao2023fsdp,jacobs2023ulysses,%
rajbhandari2022deepspeedmoe}.  What none of them bounds is the four live sets
of \cref{tab:main:operators}, whose heights follow the workload and the model
dimensions rather than the sharding degrees.

\begin{table}[t]
\centering
\small
\caption{The four live sets of tensors that this paper targets to reduce peak memory by making large HBM materializations into streams. 
For MoE layers, $R_d$ is the routes received by destination $d$, $E_p$ the expert-parallel degree, $k$ the routing degree,
$\ceff$ the effective per-chunk token budget at and $q=k\ceff$ the
routes it emits. For projection layer, $P$ the tensor-parallel group size, For SCO, $\mathcal S$ the offloaded layer
set, $N_{\max}$ the per-rank token ceiling, and $\beta$ the optimizer bucket
size.  
}
\label{tab:main:operators}
\setlength{\tabcolsep}{3.5pt}
\begin{tabular}{@{}L{0.19\textwidth}L{0.26\textwidth}L{0.18\textwidth}L{0.25\textwidth}@{}}
\toprule
Live set & Unbounded size, and what sets it & Stream axis & Bounded working set \\
\midrule
Expert dispatch &
  $2R_dHb$; realized routing matrix &
  token chunks &
  $2E_pqHb$ \\
Vocabulary projection &
  $\mathcal O(NVb)$ logits; tokens $\times$ vocabulary &
  vocabulary blocks &
  $\mathcal O(NVb/P)$ \\
Checkpoint boundaries &
  $\sum_{\ell\in\mathcal S}N_\ell Hb$ in HBM; depth $\times$ length &
  decoder depth &
  $\leq2N_{\max}Hb$ in HBM \\
Offloaded AdamW &
  serial host update of $12\Theta/W$ bytes &
  parameter buckets &
  $\mathcal O(\beta)$ in HBM \\
\bottomrule
\end{tabular}
\end{table}

\paragraph{What each operator has to deliver.}
The four rows scale differently and need different schedules, but we hold each
operator to the same requirements and check them one at a time in
\cref{sec:main:method}. Each operator has to return the
forward values and the gradients of the implementation it replaces. It may only change the
order of computation but the results must be the same.  And its memorybound has to survive checkpoint
recomputation, which rules out retaining anything sized by the routing matrix.
As a checkpointed step reaches its peak inside backward, a bound that holds
only in the original forward pass is not worth much.
These four rows are the terms of the per-rank budget we target, collected in \cref{eq:main:total-peak}; the remaining consumers of device memory are handled by established components.

\section{Four Bounded-Streaming Operators}\label{sec:main:method}

Each subsection gives the mechanism, the bound it delivers, and a benchmark that isolates the live set the operator governs.  Benchmarks run on one eight-H200 node and hold the input fixed across the systems compared; configurations are in \cref{app:setup}. Composition follows in \cref{sec:main:experiments}.

\subsection{\lleptwo: Receiver-Bounded Expert Dispatch}
\label{sec:main:llep}

\begin{figure}[t]
\centering
\begin{tikzpicture}[
  oomlbl/.style={font=\fontsize{5.4}{6}\selectfont, text=black!62,
                 rotate=90, anchor=west}
]
\begin{groupplot}[
  group style={group size=2 by 1, horizontal sep=1.30cm},
  width=0.46\textwidth, height=0.20\textwidth,
  ybar=0.5pt, /tikz/bar width=3.6pt,
  tick label style={font=\scriptsize},
  label style={font=\scriptsize},
  title style={font=\scriptsize\bfseries, yshift=-1.2mm},
  ymajorgrids=true, grid style={dashed,gray!25},
  axis line style={black!45}, tick style={black!45},
  xtick={1,2,3,4,5},
  xticklabels={bal.,30/16,50/16,80/16,95/16},
  xticklabel style={font=\fontsize{6}{7}\selectfont, rotate=42, anchor=east},
  xmin=0.45, xmax=5.55,
  legend style={font=\fontsize{6}{7}\selectfont, draw=none, fill=none,
                at={(1.10,-0.70)}, anchor=north, legend columns=3,
                column sep=8pt, inner sep=1pt},
  legend cell align=left
]
\nextgroupplot[title={(a) Peak allocated memory (GiB)},
  ymin=0, ymax=62, ytick={0,20,40,60}]
\addplot[draw=baselinegray, fill=baselinegray!45] coordinates {(1,52.660)};
\addlegendentry{standard EP}
\addplot[draw=offloadorange, fill=offloadorange!45] coordinates {
  (1,52.660) (2,52.906) (3,52.906) (4,52.988) (5,52.988)};
\addlegendentry{LLEP}
\addplot[draw=routeblue, fill=routeblue!55] coordinates {
  (1,21.430) (2,22.772) (3,22.773) (4,22.861) (5,22.855)};
\addlegendentry{\lleptwo}
\node[oomlbl] at ([xshift=-4.1pt]axis cs:2,1.5) {OOM};
\node[oomlbl] at ([xshift=-4.1pt]axis cs:3,1.5) {OOM};
\node[oomlbl] at ([xshift=-4.1pt]axis cs:4,1.5) {OOM};
\node[oomlbl] at ([xshift=-4.1pt]axis cs:5,1.5) {OOM};

\nextgroupplot[title={(b) Forward latency (ms)},
  ymin=0, ymax=215, ytick={0,50,100,150,200}]
\addplot[draw=baselinegray, fill=baselinegray!45] coordinates {(1,177.82)};
\addplot[draw=offloadorange, fill=offloadorange!45] coordinates {
  (1,177.82) (2,182.97) (3,186.07) (4,189.14) (5,182.13)};
\addplot[draw=routeblue, fill=routeblue!55] coordinates {
  (1,161.00) (2,176.62) (3,182.64) (4,180.30) (5,178.57)};
\node[oomlbl] at ([xshift=-4.1pt]axis cs:2,5) {OOM};
\node[oomlbl] at ([xshift=-4.1pt]axis cs:3,5) {OOM};
\node[oomlbl] at ([xshift=-4.1pt]axis cs:4,5) {OOM};
\node[oomlbl] at ([xshift=-4.1pt]axis cs:5,5) {OOM};
\end{groupplot}
\end{tikzpicture}

\vspace{0.4em}
\small
\setlength{\tabcolsep}{6pt}
\begin{tabular}{@{}L{0.28\textwidth} R{0.13\textwidth} R{0.16\textwidth} R{0.12\textwidth} R{0.15\textwidth}@{}}
\toprule
Shape & LLEP (GiB) & \lleptwo (GiB) & Peak saved & Speedup vs.\ LLEP \\
\midrule
65K tokens/rank, $H{=}7168$, top-8
  & 52.7--53.0 & \textbf{21.4--22.9} & \textbf{56.9--59.3\%}
  & $1.01$--$1.10\times$ \\
\bottomrule
\end{tabular}
\caption{Expert dispatch as routing skew grows, at 65,536 tokens per rank, $H=7168$, $I=2048$, 128 experts, top-$8$, and $c=6554$ ($K=10$). A profile skew-level label gives the degree of routing imbalance (such as 95/16, which means 95\% tokens routed to hottest 16 experts). \cref{app:llep-profiles} defines the family.  At this shape standard EP runs out of memory as soon as the router is severely skewed. PipelinedLLEP saves up to 59\% peak memory without significant slowdown. More details are provided in Appendix \cref{app:llep-profiles,tab:llep-sweeps-full,tab:llep-length}.
}
\label{fig:main:llep-sweep}
\end{figure}
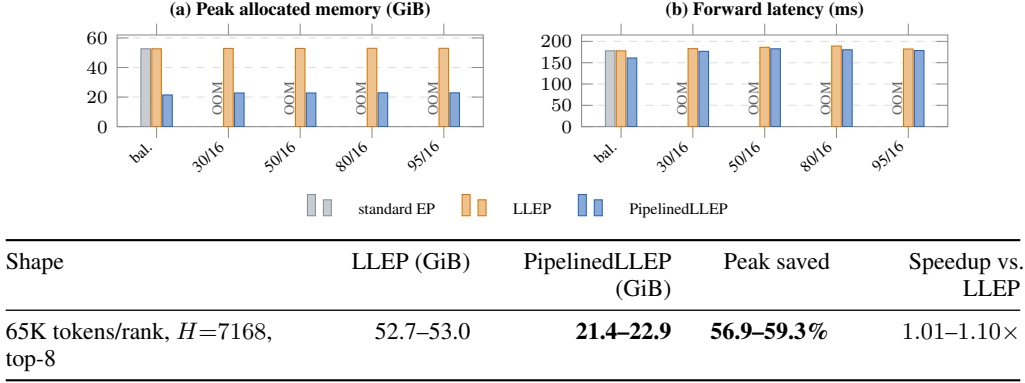

Under expert parallelism each rank owns only some of the experts, so a token routed to one of them has to travel to the rank that owns it, its \emph{destination} $d$.  The number of routes $R_d$ to $d$ determines the size of its dispatch buffers \citep{gshard_lepikhin2020,rajbhandari2022deepspeedmoe}. That count grows with three things: the batch size, the routing top-k $k$, and routing imbalance. LLEP removes the third of these by moving a popular expert's overflow onto ranks that would otherwise be idle \citep{nguyen2026llep}. The first two remain: LLEP still holds the rank's entire routed batch as the all-to-all materializes that batch in one tensor (\cref{fig:main:llep-sweep}).

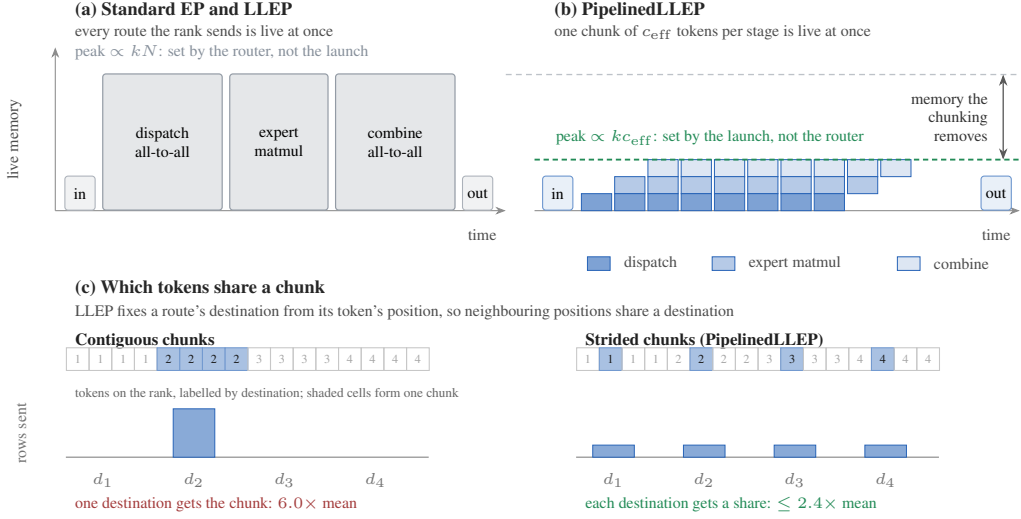
\begin{figure}[t]
  \centering
\begin{tikzpicture}[
  x=1cm, y=1cm,
  tile/.style={line width=0.3pt, draw=routeblue},
  bigbox/.style={rounded corners=1.2pt, line width=0.45pt,
                 fill=baselinegray!22, draw=baselinegray!90,
                 fglbl, align=center, inner sep=0pt},
  tokbox/.style={rounded corners=1.2pt, line width=0.4pt, fglbl,
                 align=center, inner sep=0pt},
  caplbl/.style={fglbl, text=black!70},
  sublbl/.style={font=\fontsize{6.4}{7.4}\selectfont\bfseries, text=black!88},
  cellbase/.style={line width=0.4pt, minimum size=0.30cm, inner sep=0pt,
                   font=\fontsize{4.6}{5}\selectfont, anchor=south west},
  celloff/.style={cellbase, draw=black!22, text=black!35},
  cellon/.style={cellbase, draw=routeblue!65, fill=routeblue!40, text=black!90},
]

\def\FullH{1.80}   
\def\LaneA{0.225}  
\def\LaneB{0.45}
\def\LaneC{0.675}
\def\TokH{0.45}    

\draw[fgaxis, -{Stealth[length=1.4mm]}] (0.66,0) -- (0.66,2.16);
\node[caplbl, rotate=90, anchor=south] at (0.34,0.95) {live memory};

\node[fgtitle, anchor=north west] at (0.80,2.88) {(a) Standard EP and LLEP};
\node[caplbl, anchor=north west] at (0.80,2.58)
  {every route the rank sends is live at once};

\draw[fgaxis, -{Stealth[length=1.4mm]}] (0.66,0) -- (6.62,0);
\node[caplbl, anchor=north east] at (6.62,-0.12) {time};

\node[tokbox, fill=baselinegray!12, draw=baselinegray!70,
      minimum width=0.40cm, minimum height=\TokH cm, anchor=south west]
  at (0.78,0) {in};
\node[bigbox, minimum width=1.58cm, minimum height=\FullH cm, anchor=south west]
  at (1.28,0) {dispatch\\all-to-all};
\node[bigbox, minimum width=1.30cm, minimum height=\FullH cm, anchor=south west]
  at (2.96,0) {expert\\matmul};
\node[bigbox, minimum width=1.58cm, minimum height=\FullH cm, anchor=south west]
  at (4.36,0) {combine\\all-to-all};
\node[tokbox, fill=baselinegray!12, draw=baselinegray!70,
      minimum width=0.40cm, minimum height=\TokH cm, anchor=south west]
  at (6.04,0) {out};

\node[caplbl, text=baselinegray, anchor=south west] at (0.80,\FullH+0.07)
  {peak $\propto kN$: set by the router, not the launch};

\node[fgtitle, anchor=north west] at (7.14,2.88) {(b) \lleptwo};
\node[caplbl, anchor=north west] at (7.14,2.58)
  {one chunk of $\ceff$ tokens per stage is live at once};

\draw[fgaxis, -{Stealth[length=1.4mm]}] (7.00,0) -- (13.42,0);
\node[caplbl, anchor=north east] at (13.42,-0.12) {time};

\node[tokbox, fill=routeblue!10, draw=routeblue!70,
      minimum width=0.40cm, minimum height=\TokH cm, anchor=south west]
  at (7.10,0) {in};

\foreach \i in {1,...,8}{
  \pgfmathsetmacro{\xa}{7.62+(\i-1)*0.44}
  \pgfmathsetmacro{\xb}{7.62+\i*0.44}
  \pgfmathsetmacro{\xc}{7.62+(\i+1)*0.44}
  \filldraw[tile, fill=routeblue!60] (\xa,0)      rectangle +(0.40,\LaneA);
  \filldraw[tile, fill=routeblue!34] (\xb,\LaneA) rectangle +(0.40,\LaneA);
  \filldraw[tile, fill=routeblue!16] (\xc,\LaneB) rectangle +(0.40,\LaneA);
}

\node[tokbox, fill=routeblue!10, draw=routeblue!70,
      minimum width=0.40cm, minimum height=\TokH cm, anchor=south west]
  at (12.90,0) {out};

\draw[fgbound] (7.00,\LaneC) -- (13.42,\LaneC);
\node[caplbl, text=successgreen, anchor=south west] at (7.16,\LaneC+0.07)
  {peak $\propto k\ceff$: set by the launch, not the router};

\draw[draw=baselinegray!55, line width=0.5pt, dash pattern=on 2.2pt off 1.6pt]
  (6.62,\FullH) -- (13.42,\FullH);
\draw[{Stealth[length=1.5mm]}-{Stealth[length=1.5mm]},
      draw=black!65, line width=0.5pt] (13.20,\FullH) -- (13.20,\LaneC);
\node[caplbl, anchor=east, align=right, text=black!80] at (13.12,1.24)
  {memory the\\chunking\\removes};

\filldraw[tile, fill=routeblue!60] (7.70,-0.79) rectangle +(0.30,0.18);
\node[caplbl, anchor=west] at (8.06,-0.70) {dispatch};
\filldraw[tile, fill=routeblue!34] (9.35,-0.79) rectangle +(0.30,0.18);
\node[caplbl, anchor=west] at (9.71,-0.70) {expert matmul};
\filldraw[tile, fill=routeblue!16] (11.80,-0.79) rectangle +(0.30,0.18);
\node[caplbl, anchor=west] at (12.16,-0.70) {combine};

\begin{scope}[yshift=0.26cm]
\node[fgtitle, anchor=north west] at (0.80,-1.05)
  {(c) Which tokens share a chunk};
\node[caplbl, anchor=north west] at (0.80,-1.39)
  {LLEP fixes a route's destination from its token's position, so neighbouring
   positions share a destination};

\node[caplbl, text=black!55, rotate=90, anchor=south] at (0.40,-3.20) {rows sent};

\node[sublbl, anchor=north west] at (0.80,-1.76) {Contiguous chunks};
\foreach \j/\dg/\st in {0/1/celloff, 1/1/celloff, 2/1/celloff, 3/1/celloff,
                        4/2/cellon,  5/2/cellon,  6/2/cellon,  7/2/cellon,
                        8/3/celloff, 9/3/celloff, 10/3/celloff, 11/3/celloff,
                        12/4/celloff, 13/4/celloff, 14/4/celloff, 15/4/celloff}{
  \pgfmathsetmacro{\cx}{0.80+\j*0.30}
  \node[\st] at (\cx,-2.38) {\dg};
}
\node[caplbl, text=black!60, anchor=north west, font=\fontsize{5}{6}\selectfont]
  at (0.80,-2.48)
  {tokens on the rank, labelled by destination; shaded cells form one chunk};
\draw[draw=black!40, line width=0.5pt] (0.80,-3.52) -- (5.60,-3.52);
\filldraw[tile, fill=routeblue!55] (2.22,-3.52) rectangle +(0.55,0.64);
\foreach \j in {0,...,3}{
  \pgfmathsetmacro{\bx}{1.02+\j*1.20+0.275}
  \pgfmathtruncatemacro{\dn}{\j+1}
  \node[caplbl, text=black!55, anchor=north] at (\bx,-3.60) {$d_{\dn}$};
}
\node[caplbl, text=warnred, anchor=north west] at (0.80,-3.92)
  {one destination gets the chunk: $6.0\times$ mean};

\node[sublbl, anchor=north west] at (7.55,-1.76)
  {Strided chunks (\lleptwo)};
\foreach \j/\dg/\st in {0/1/celloff, 1/1/cellon,  2/1/celloff, 3/1/celloff,
                        4/2/celloff, 5/2/cellon,  6/2/celloff, 7/2/celloff,
                        8/3/celloff, 9/3/cellon,  10/3/celloff, 11/3/celloff,
                        12/4/celloff, 13/4/cellon, 14/4/celloff, 15/4/celloff}{
  \pgfmathsetmacro{\cx}{7.55+\j*0.30}
  \node[\st] at (\cx,-2.38) {\dg};
}
\draw[draw=black!40, line width=0.5pt] (7.55,-3.52) -- (12.35,-3.52);
\foreach \j in {0,...,3}{
  \pgfmathsetmacro{\bx}{7.77+\j*1.20}
  \pgfmathsetmacro{\bc}{7.77+\j*1.20+0.275}
  \pgfmathtruncatemacro{\dn}{\j+1}
  \filldraw[tile, fill=routeblue!55] (\bx,-3.52) rectangle +(0.55,0.16);
  \node[caplbl, text=black!55, anchor=north] at (\bc,-3.60) {$d_{\dn}$};
}
\node[caplbl, text=successgreen, anchor=north west] at (7.55,-3.92)
  {each destination gets a share: $\leq2.4\times$ mean};
\end{scope}

\end{tikzpicture}
  \caption{\lleptwo limits expert dispatch by capping how many tokens each rank puts into a chunk. \textbf{(a)} Dispatching the whole batch at once causes peak memory to spike. \textbf{(b)} With $K$ chunks of $\ceff$ tokens, a stage holds $1/K$ as much and the stages overlap, so peak memory stays under the limit of \cref{eq:main:llep-bound}. 
  \textbf{(c)} LLEP rebalances loads by arranging token destinations into contiguous indices. But when the tokens are dispatched in chunks, a chunk of consecutive positions sends almost all of its routes to the same rank, reintroducing routing imbalance. \lleptwo instead selects tokens for each chunk in strided order, spreading each chunk's routes over the destinations, which balances the load for each chunk.
  At a 65K tokens batch, consecutive chunks causes routing peak of $6.0\times$ the average, while strided chunks stay within $2.4\times$, which is worth up to $1.35\times$ in speed and up to $1.37\gib$ of peak memory (\cref{tab:llep-membership}).}
  \label{fig:main:llep-diagram}
\end{figure}

\lleptwo instead splits the batch into chunks, using a maximum token budget $c$, and processes the chunks in a pipeline that overlaps communication with computation.  The budget applies to each sending rank separately: a rank puts at most $c$ of its own tokens into any one chunk. The number of chunks $K$ is computed as:
\begin{equation}
  K=\min\!\left(\left\lceil N/c\right\rceil,\,K_{\max}\right), \qquad \ceff=\left\lceil N/K\right\rceil, \qquad R_d^{(i)}\le E_pk\ceff ,
  \label{eq:main:llep-bound}
\end{equation}
where $N$ is the number of tokens on a rank, $k$ is the router's top-$k$, $E_p$ is the expert-parallel degree, and $K_{\max}$ is a predetermined limit on the number of chunks.  $K$ is the number of chunks, $\ceff$ is how many tokens a rank actually puts in one chunk, and $R_d^{(i)}$ is the number of routed tokens that destination $d$ receives in chunk $i$.
Each of the $E_p$ ranks puts at most $\ceff$ tokens into chunk $i$, and each token goes to $k$ experts, so one rank sends at most $k\ceff$ routes in that chunk no matter which experts its tokens chose. Summing over the $E_p$ senders produces $E_pk\ceff$. No routed token is dropped to make this hold.
\cref{eq:main:llep-bound} contains only $E_p$, $k$, and $\ceff$, and none of them depend on what the router does (\cref{fig:main:llep-diagram}a,b).  A skewed router still sends more routes to one destination than to another, but it cannot push any buffer past this limit. When $N\le cK_{\max}$, the chunk limit is never reached and $\ceff=c$, so the buffers take at most $2E_pkcHb$ bytes together at hidden width $H$ and $b$ bytes per value.  When $N>cK_{\max}$, the number of chunks stops at $K_{\max}$ and each buffer is capped at a fixed $1/K_{\max}$. Either way the limit is known beforehand. Deriving $K$ from $c$ allows the system to dynamically handle micro-batch length variations inflight without facing OOM.

\paragraph{Why overlap alone is not enough.}
The $K$ chunks run through a three-stage pipeline: dispatch, then the grouped expert matmul \citep{megablocks}, then combine.  While chunk $i$ is on the tensor cores, the dispatch of chunk $i{+}1$ and the combine of chunk $i{-}1$ use the interconnect.  Other systems build this same pipeline, but they choose the number of chunks to make the iteration fast \citep{fastermoe,pipemoe,tutel,schemoe}; we choose it to cap memory instead, which is a different criterion and gives a different guarantee.  Overlap on its own also does not lower the peak during training, because a loop that calls the expert layer once per chunk keeps every chunk's autograd graph alive until the layer's backward pass runs.  \lleptwo therefore wraps each chunk's expert matmul in a reentrant gradient checkpoint, nested inside the decoder layer's non-reentrant gradient checkpoint \citep{chen2016sublinear,pytorchcheckpointwrapper}.  
The inner checkpoint frees one chunk's intermediate tensors before the next chunk allocates its own, both forward and backward. \Cref{app:llep-backward} gives the details and the gradient accounting.

\paragraph{Which tokens go in which chunk.}
LLEP picks the rank that handles a route from the position of its token in the expert's global token list.  A chunk of consecutive positions inherits that grouping, so a chunk covering the relocated tail of a popular expert sends everything it has to a single helper rank, even when the layer's plan is balanced overall.  \lleptwo therefore gives chunk $i$ the positions $i,i{+}K,i{+}2K,\dots$ (\cref{fig:main:llep-diagram}c), which spreads each chunk's routes over the destinations. 
Nothing else changes: the same tokens are sent, the relocation plan is the same, and every route goes to the same destination as before. 
Striding runs $1.03$--$1.35\times$ faster and saves up to $1.37\gib$ of peak memory (\cref{tab:llep-membership}).


\paragraph{Isolated benchmark.}
As shown in \cref{fig:main:llep-sweep}, \lleptwo saves $56.9$--$59.3\%$ over LLEP at $1.01$--$1.10\times$ its speed; the saving grows with the unpartitioned live set.  Peak memory is deterministic in the configuration and verified empirically. Standard expert parallelism is the faster of the three only where there is no skew to correct and no concentration to cap.

\subsection{\ringtp: Exact Vocabulary Projection over Distinct Batches}
\label{sec:main:ringtp}

A token's cross-entropy loss does not need its whole logit row, only a running
maximum, exponential sum, and the logit of its target.  Those three
scalars are the same size regardless of vocabulary size, so
the vocabulary can be visited one block at a time and the logit tensor never
has to exist. Fused cross-entropy kernels use this on a single device
\citep{wijmans2025cce,hsu2024liger}; the difficulty here is that the batch and
the projection weight are \emph{both} sharded across ranks.

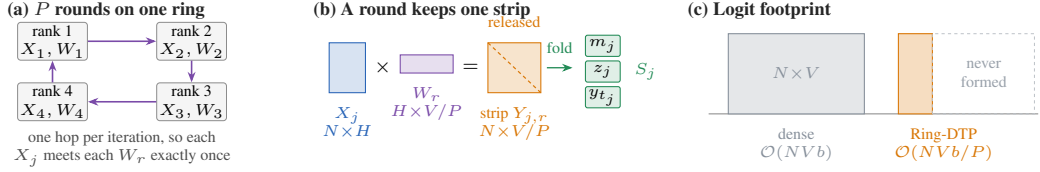
\begin{figure}[t]
  \centering
\begin{tikzpicture}[
  rank/.style={draw=black!45, line width=0.35pt, rounded corners=1.2pt,
               fill=black!4, inner sep=1.0pt, fglbl, align=center},
  hop/.style={draw=vocabpurple, line width=0.6pt,
              -{Stealth[length=1.4mm]}},
  statebox/.style={draw=successgreen, line width=0.4pt, fill=successgreen!14,
                   rounded corners=0.8pt, minimum width=0.46cm,
                   minimum height=0.28cm, inner sep=0.4pt, fglbl}
]

\node[fgtitle, anchor=north west] at (-0.10,0.82)
  {(a) $P$ rounds on one ring};

\node[rank] (r1) at (0.62,0.18) {rank 1\\$X_1,W_1$};
\node[rank] (r2) at (2.46,0.18) {rank 2\\$X_2,W_2$};
\node[rank] (r3) at (2.46,-0.62) {rank 3\\$X_3,W_3$};
\node[rank] (r4) at (0.62,-0.62) {rank 4\\$X_4,W_4$};

\draw[hop] (r1) -- (r2);
\draw[hop] (r2) -- (r3);
\draw[hop] (r3) -- (r4);
\draw[hop] (r4) -- (r1);

\node[fglbl, text=black!70, anchor=north, align=center] at (1.54,-0.88)
  {one hop per iteration, so each\\$X_j$ meets each $W_r$ exactly once};

\node[fgtitle, anchor=north west] at (3.95,0.82)
  {(b) A round keeps one strip};

\draw[draw=routeblue, line width=0.4pt, fill=routeblue!18]
  (4.28,-0.50) rectangle (4.76,0.16);
\node[fglbl, text=routeblue, anchor=north, align=center] at (4.52,-0.56)
  {$X_j$\\$N{\times}H$};

\node[font=\scriptsize] at (4.99,-0.17) {$\times$};

\draw[draw=vocabpurple, line width=0.4pt, fill=vocabpurple!18]
  (5.22,-0.24) rectangle (5.92,0.00);
\node[fglbl, text=vocabpurple, anchor=north, align=center] at (5.57,-0.30)
  {$W_r$\\$H{\times}V/P$};

\node[font=\scriptsize] at (6.15,-0.17) {$=$};

\draw[draw=offloadorange, line width=0.4pt, fill=offloadorange!20]
  (6.38,-0.50) rectangle (7.08,0.16);
\draw[draw=offloadorange, line width=0.45pt, dash pattern=on 1.4pt off 1.2pt]
  (6.38,0.16) -- (7.08,-0.50);
\node[fglbl, text=offloadorange, anchor=north, align=center] at (6.73,-0.56)
  {strip $Y_{j,r}$\\$N{\times}V/P$};
\node[fglbl, text=offloadorange, anchor=south] at (6.73,0.24) {released};

\draw[-{Stealth[length=1.4mm]}, draw=successgreen, line width=0.5pt]
  (7.16,-0.17) -- (7.50,-0.17);
\node[fglbl, text=successgreen, anchor=south] at (7.33,-0.10) {fold};

\node[statebox] at (7.90,0.12) {$m_j$};
\node[statebox] at (7.90,-0.22) {$z_j$};
\node[statebox] at (7.90,-0.56) {$y_{t_j}$};
\node[fglbl, text=successgreen, anchor=west] at (8.20,-0.22) {$S_j$};

\node[fgtitle, anchor=north west] at (8.90,0.82)
  {(c) Logit footprint};

\draw[fgaxis] (9.30,-0.78) -- (13.66,-0.78);

\fill[baselinegray!22] (9.56,-0.78) rectangle (11.36,0.28);
\draw[draw=baselinegray, line width=0.4pt]
  (9.56,-0.78) rectangle (11.36,0.28);
\node[fglbl, text=baselinegray] at (10.46,-0.25) {$N{\times}V$};
\node[fglbl, text=baselinegray, anchor=north, align=center] at (10.46,-0.84)
  {dense\\$\mathcal O(NVb)$};

\fill[offloadorange!25] (11.81,-0.78) rectangle (12.26,0.28);
\draw[draw=offloadorange, line width=0.4pt]
  (11.81,-0.78) rectangle (12.26,0.28);
\draw[draw=baselinegray!55, line width=0.3pt,
      dash pattern=on 1.2pt off 1.2pt]
  (12.26,-0.78) rectangle (13.61,0.28);
\node[fglbl, text=baselinegray!85, align=center, anchor=center] at (12.94,-0.25)
  {never\\formed};
\node[fglbl, text=offloadorange, anchor=north, align=center] at (12.41,-0.84)
  {\ringtp\\$\mathcal O(NVb/P)$};

\end{tikzpicture}
  \caption{\ringtp evaluates the vocabulary projection without materializing
  logits.  \textbf{(a)} Ranks hold distinct local batches and distinct
  vocabulary shards; one hop per iteration puts every batch with every
  shard exactly once, drawn for $P=4$.  \textbf{(b)} A round forms one strip
  $Y_{j,r}$, folds its normalizer and target logit into the three-scalar
  per-token state $S_j$, and releases the strip.  \textbf{(c)} Only one strip
  of $V/P$ columns is ever live, so the logit footprint falls by $P$ and the
  $N{\times}V$ tensor is never formed.}
  \label{fig:main:ringtp-diagram}
\end{figure}

Rank $r$ holds a distinct local batch $X_r\in\mathbb R^{N\times H}$ with
targets $t_r$ and owns the vocabulary interval $\mathcal V_r$ of size $V/P$, so
$W_{\mathrm{voc}}\in\mathbb R^{H\times V}$ is column-sharded as
$W_r:=W_{\mathrm{voc}}[:,\mathcal V_r]$.  Megatron's vocabulary-parallel uses the same weight partition but requires every rank to hold the
same $X$ \citep{shoeybi2019megatron}, which reduce the effective batch size and long-context scalability.
By using a data-parallel-like layout, \ringtp
preserves the distinct $X_r$ and arranges $P$ ring-like rounds of data transfer and computation. Over those rounds, every batch
$X_j$ is co-located with every shard $W_r$ once
(\cref{fig:main:ringtp-diagram}).
A round forms only the logit strip
$Y_{j,r}=X_jW_r\in\mathbb R^{N\times V/P}$, folds its partial normalizer and
target logit into the running per-token state $S_j=(m_j,z_j,y_{t_j})$, and
releases the strip. Peak logit memory is then
\begin{equation}
  M_{\mathrm{logit}}^{\ringtp}=\mathcal O(NVb/P),
  \label{eq:main:ringtp-bound}
\end{equation}
where $b$ is the number of bytes per value.
For one token, the logits $y$ of each strip update the running state in the
standard online-softmax form \citep{milakov2018online,dao2022flashattention}:
\begin{equation}
 m'=\max(m,\max_v y_v),\qquad
 z'=e^{m-m'}z+\sum_v e^{y_v-m'} .
 \label{eq:main:ringtp-online}
\end{equation}
After all $P$ rounds the exact negative log-likelihood is $m+\log z-y_t$.
Backward recomputes the gradients one strip at a time from the saved normalizer via the same ring, so neither pass stores a full logit tensor.

\paragraph{Which tensor travels, and at what price.}
Either the activations $X_j$ or the weight shard $W_r$ can move. \ringtp chooses the cheapest option dynamically. The move-weights schedule (\cref{fig:ringtp-move-weights}) keeps $(X_j,t_j,S_j)$ on its rank and sends $W_r$ around the ring, so forward needs no return hop, unlike a ring that can only move one of the two tensors \citep{liu2024ringattention}.  Bytes sent per hop are $\mathcal O(NH)$ if activations move and $\mathcal O(HV/P)$ if weights move. We move weights exactly when $N>V/P$.  Backward replays the same rounds, and the gradient of the moving tensor travels with it.  Sharding also cuts persistent projection storage and its backward workspace by $P$.  The cost is $P{-}1$ sequential hops per pass. We therefore choose the smallest $P$ for which the $NV/P$ strip fits.  Target-column ownership and gradient derivations are in \cref{app:ringtp}.

\begin{table}[t]
\centering
\small
\caption{Vocabulary projection at $P=8$ on a distinct local batch of $N$ per
GPU with $H=7{,}168$ and $V=200{,}000$ in FP32, measured over hidden states to
target-token log-probabilities and backward.  The standard reference
materializes full logits and log-softmax.  Auto-mode picks move-weights when
$N>V/P$, so the two rows cover one branch each: $N=16{,}384$ falls below
$V/P=25{,}000$ and $N=32{,}768$ exceeds it. Peak and latency come from one
measured pass.  Per-tensor
detail is in \cref{tab:ringtp-memory-benchmark,tab:ringtp-forward-benchmark}.}
\label{tab:main:ringtp}
\setlength{\tabcolsep}{5pt}
\begin{tabular}{@{}rlrrrrrr@{}}
\toprule
& & \multicolumn{3}{c}{Forward+backward peak (GiB)}
& \multicolumn{3}{c}{Latency (ms)} \\
\cmidrule(lr){3-5}\cmidrule(lr){6-8}
$N$ & Schedule & Standard & \ringtp & Saved & Standard & \ringtp & Cost \\
\midrule
16,384 & move-activations & 42.462 & 7.322 & 82.8\% & 1001.2 & 1052.2 & $+5.1\%$ \\
32,768 & move-weights & 79.521 & \textbf{10.622} & \textbf{86.6\%} & 2065.2 & 2155.8 & $+4.4\%$ \\
\bottomrule
\end{tabular}
\end{table}

\paragraph{Isolated benchmark.}
\Cref{tab:main:ringtp} measures both branches at $P=8$, with the local batch $N$ chosen
so that auto-mode moves activations in one row and weights in the other. In both cases, the peak falls by $82.8\%$ and $86.6\%$
for at most $5.1\%$ more time.  Doubling $N$ nearly doubles the standard
baseline, from $42.5$ to $79.5\gib$, while \ringtp moves from $7.3$ to
$10.6\gib$, so the saving widens as context grows. That is what lets the
vocabulary head stay on device at 1M context length, where the logit
tensor alone would exceed HBM. The move-weights branch scales best with context length, as its per-hop payload $\mathcal O(HV/P)$ does not grow with token count.

\subsection{\selectoffload: Exact Gradient Checkpointing Boundary Offload}
\label{sec:main:offload}

Under gradient checkpointing, the layer input $h_\ell\in\mathbb R^{N_\ell\times H}$ stays on the device from the forward pass until that layer is recomputed in backward. This retained input is the checkpoint boundary.

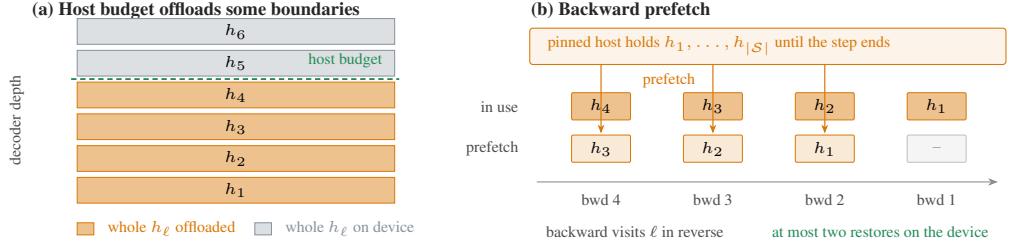
\begin{figure}[t]
  \centering
\begin{tikzpicture}[
  x=1cm, y=1cm,
  caplbl/.style={fglbl, text=black!70},
  bndoff/.style={draw=offloadorange, fill=offloadorange!45, line width=0.35pt},
  bndon/.style={draw=baselinegray, fill=baselinegray!28, line width=0.35pt},
  slot/.style={rounded corners=0.6pt, line width=0.35pt, inner sep=0pt,
               fglbl, align=center, minimum height=0.36cm, minimum width=0.78cm,
               anchor=center},
  hostbox/.style={draw=offloadorange, line width=0.4pt, rounded corners=1.2pt,
                  fill=offloadorange!8}
]

\begin{scope}
\node[fgtitle, anchor=north west] at (0.00,2.78)
  {(a) Host budget offloads some boundaries};

\foreach \scoi/\scolab/\scostyle in {
  1/h_1/bndoff,
  2/h_2/bndoff,
  3/h_3/bndoff,
  4/h_4/bndoff,
  5/h_5/bndon,
  6/h_6/bndon}{
  \pgfmathsetmacro{\scoy}{(\scoi-1)*0.42}
  \filldraw[\scostyle] (0.72,\scoy) rectangle ++(4.20,0.34);
  \node[fglbl] at (2.82,\scoy+0.17) {$\scolab$};
}

\node[caplbl, rotate=90, anchor=south] at (0.14,1.10) {decoder depth};

\draw[fgbound] (0.64,1.64) -- (4.98,1.64);
\node[fglbl, text=successgreen, anchor=south east] at (4.88,1.67)
  {host budget};

\filldraw[bndoff] (0.72,-0.42) rectangle ++(0.22,0.16);
\node[fglbl, text=offloadorange, anchor=west] at (1.00,-0.34)
  {whole $h_\ell$ offloaded};
\filldraw[bndon] (3.07,-0.42) rectangle ++(0.22,0.16);
\node[fglbl, text=baselinegray, anchor=west] at (3.35,-0.34)
  {whole $h_\ell$ on device};
\end{scope}

\begin{scope}[xshift=6.60cm]
\node[fgtitle, anchor=north west] at (0.00,2.78)
  {(b) Backward prefetch};

\node[hostbox, minimum width=6.30cm, minimum height=0.48cm,
      anchor=north west] at (0.10,2.32) {};
\node[fglbl, text=offloadorange, anchor=west] at (0.22,2.08)
  {pinned host holds $h_1,\ldots,h_{|\mathcal S|}$ until the step ends};

\node[caplbl, anchor=east] at (0.08,1.28) {in use};
\node[caplbl, anchor=east] at (0.08,0.72) {prefetch};

\foreach \scok/\scouse/\scopf in {0/h_4/h_3, 1/h_3/h_2, 2/h_2/h_1}{
  \pgfmathsetmacro{\scocx}{1.05+\scok*1.48}
  \node[slot, bndoff] at (\scocx,1.28) {$\scouse$};
  \node[slot, draw=offloadorange, fill=offloadorange!16] at (\scocx,0.72)
    {$\scopf$};
  \draw[-{Stealth[length=1.2mm]}, draw=offloadorange, line width=0.45pt]
    (\scocx,1.84) -- (\scocx,0.92);
}

\node[slot, bndoff] at (5.49,1.28) {$h_1$};
\node[slot, draw=black!25, fill=black!4, text=black!40] at (5.49,0.72) {--};

\node[fglbl, text=offloadorange, anchor=west] at (1.48,1.62) {prefetch};

\draw[fgaxis, -{Stealth[length=1.3mm]}] (0.20,0.28) -- (6.30,0.28);
\node[caplbl, anchor=north] at (1.05,0.24) {bwd 4};
\node[caplbl, anchor=north] at (2.53,0.24) {bwd 3};
\node[caplbl, anchor=north] at (4.01,0.24) {bwd 2};
\node[caplbl, anchor=north] at (5.49,0.24) {bwd 1};
\node[caplbl, anchor=north west] at (0.20,-0.18)
  {backward visits $\ell$ in reverse};
\node[fglbl, text=successgreen, anchor=north east] at (6.30,-0.18)
  {at most two restores on the device};
\end{scope}

\end{tikzpicture}
  \caption{\selectoffload offloads checkpoint boundaries up to a host budget and prefetches them during backward. \textbf{(a)} Each bar is one layer's checkpoint boundary $h_\ell$. Orange bars fit the host budget and are copied entirely to pinned memory; gray bars stay on the device. \textbf{(b)} During backward, while $h_\ell$ is in use, $h_{\ell-1}$ is copied back asynchronously.}
  \label{fig:main:sco-diagram}
\end{figure}

\selectoffload moves a subset of those boundaries to pinned host memory. It walks the checkpointed layers in forward order and offloads each boundary that still fits in the host budget; those layers form the set $\mathcal S$ (\cref{fig:main:sco-diagram}a). Each selected boundary is copied to the host asynchronously during forward, and the device copy is then freed. Layers that do not fit keep their boundaries on the device.

Backward visits layers in reverse. While layer $\ell$ recomputes from $h_\ell$, the next boundary $h_{\ell-1}$ is copied back from the host on a separate stream (\cref{fig:main:sco-diagram}b). At most two restored boundaries are therefore live on the device at once. Let $N_{\max}$ be the per-rank token ceiling. The restore working set and the host occupancy of $\mathcal S$ then satisfy
\begin{equation}
 M_{\mathrm{HBM}}^{\selectoffload}\le2N_{\max}Hb,\qquad
 M_{\mathrm{host}}\le|\mathcal S|N_{\max}Hb .
 \label{eq:main:offload-bound}
\end{equation}
The offloaded set stays in host memory until the step ends. Transfer ordering and session cleanup are in \cref{app:sco}.

\begin{table}[t]
\centering
\small
\caption{Matched \selectoffload budget sweep on gpt-oss-20b \citep{gptoss}, 8xH200 GPUs,
sequence and expert parallelism of degree eight. Every policy processes a configured global batch size of 556432 tokens. Full sweeps are in \cref{tab:sco-matched-benchmark,tab:sco-capacity-benchmark}.}
\label{tab:main:sco}
\setlength{\tabcolsep}{5pt}
\begin{tabular}{@{}L{0.11\textwidth} R{0.11\textwidth} R{0.12\textwidth} R{0.12\textwidth} R{0.14\textwidth} R{0.16\textwidth}@{}}
\toprule
Host budget & Boundaries & Peak HBM (GiB) & Node RAM (GiB) & Throughput (tok/s/GPU) & Largest 10-step batch \\
\midrule
Off      & 0 / 47  & 139.790 & 402.517 & 2,641 & 557,056 \\
$8\gib$  & 21 / 47 & 133.546 & 487.990 & \textbf{2,692} & 589,824 \\
$16\gib$ & 42 / 47 & 125.677 & 573.473 & 2,687 & 622,592 \\
Full     & 47 / 47 & \textbf{123.728} & 593.410 & 2,687 & \textbf{655,360} \\
\bottomrule
\end{tabular}
\end{table}

\paragraph{Isolated benchmark.}
\Cref{tab:main:sco} shows HBM-RAM trade off. The device
peak falls monotonically with the host budget, allowing more memory headroom for other GPU tasks, where node memory rises by
approximately the same amount. Throughput moves by $1.9\%$ across settings
with no monotone trend, and while saving $17.65\%$ of HBM consumption. Measured against the largest batch that runs without an
out-of-memory error, the gain is $35.71\%$ (\cref{tab:sco-capacity-benchmark}).

\subsection{\streamadam: Bounded GPU Updates over Host CPU State}
\label{sec:main:optimizer}

\begin{figure}[t]
  \centering
\begin{tikzpicture}[
  slot/.style={rounded corners=0.5pt, line width=0.3pt,
               minimum height=0.30cm, minimum width=0.60cm,
               inner sep=0pt, anchor=west, fglbl}
]

\node[fgtitle, anchor=north west] at (-0.10,1.15)
  {(a) The offloaded update, serialized and streamed};

\node[fglbl, text=black!70, anchor=east] at (1.66,0.50) {host AdamW};
\fill[baselinegray!22] (1.76,0.35) rectangle (7.40,0.65);
\draw[draw=baselinegray, line width=0.35pt] (1.76,0.35) rectangle (7.40,0.65);
\node[fglbl, text=black!75] at (4.58,0.50)
  {one serialized update over all $\Theta/W$ parameters};

\node[fglbl, text=black!70, anchor=east] at (1.66,-0.10) {host$\to$device};
\node[fglbl, text=black!70, anchor=east] at (1.66,-0.50) {GPU AdamW};
\node[fglbl, text=black!70, anchor=east] at (1.66,-0.90) {device$\to$host};

\node[slot, fill=routeblue!35, draw=routeblue] at (1.76,-0.10) {1};
\node[slot, fill=routeblue!35, draw=routeblue] at (2.42,-0.10) {2};
\node[slot, fill=routeblue!35, draw=routeblue] at (3.08,-0.10) {3};
\node[slot, fill=routeblue!35, draw=routeblue] at (3.74,-0.10) {4};
\node[slot, fill=routeblue!35, draw=routeblue] at (4.40,-0.10) {5};

\node[slot, fill=successgreen!35, draw=successgreen] at (2.42,-0.50) {1};
\node[slot, fill=successgreen!35, draw=successgreen] at (3.08,-0.50) {2};
\node[slot, fill=successgreen!35, draw=successgreen] at (3.74,-0.50) {3};
\node[slot, fill=successgreen!35, draw=successgreen] at (4.40,-0.50) {4};
\node[slot, fill=successgreen!35, draw=successgreen] at (5.06,-0.50) {5};

\node[slot, fill=offloadorange!30, draw=offloadorange] at (3.08,-0.90) {1};
\node[slot, fill=offloadorange!30, draw=offloadorange] at (3.74,-0.90) {2};
\node[slot, fill=offloadorange!30, draw=offloadorange] at (4.40,-0.90) {3};
\node[slot, fill=offloadorange!30, draw=offloadorange] at (5.06,-0.90) {4};
\node[slot, fill=offloadorange!30, draw=offloadorange] at (5.72,-0.90) {5};

\draw[fgaxis, -{Stealth[length=1.3mm]}] (1.76,-1.24) -- (7.48,-1.24);
\node[fglbl, text=black!70, anchor=north east] at (7.48,-1.30) {time};

\draw[draw=successgreen, line width=0.4pt] (2.42,-1.12) -- (2.42,-1.20);
\draw[draw=successgreen, line width=0.4pt] (3.08,-1.12) -- (3.08,-1.20);
\node[fglbl, text=successgreen, anchor=north, align=center] at (2.75,-1.30)
  {period $\max(T_{\mathrm{H2D}},T_{\mathrm{upd}},T_{\mathrm{D2H}})$};

\node[fgtitle, anchor=north west] at (8.35,1.15)
  {(b) Device bytes per rank};

\draw[fgaxis] (8.75,-1.24) -- (13.45,-1.24);

\fill[baselinegray!22] (9.30,-1.24) rectangle (10.34,-0.80);
\draw[draw=baselinegray, line width=0.4pt] (9.30,-1.24) rectangle (10.34,-0.80);
\node[fglbl, text=black!75] at (9.82,-1.02) {$4\Theta/W$};
\fill[baselinegray!45] (9.30,-0.80) rectangle (10.34,0.52);
\draw[draw=baselinegray, line width=0.4pt] (9.30,-0.80) rectangle (10.34,0.52);
\node[fglbl, text=black!75, align=center] at (9.82,-0.14)
  {$12\Theta/W$\\state};
\node[fglbl, text=black!70, anchor=north] at (9.82,-1.32) {resident};

\fill[routeblue!25] (11.60,-1.24) rectangle (12.64,-0.80);
\draw[draw=routeblue, line width=0.4pt] (11.60,-1.24) rectangle (12.64,-0.80);
\node[fglbl, text=routeblue] at (12.12,-1.02) {$4\Theta/W$};
\fill[successgreen!35] (11.60,-0.80) rectangle (12.64,-0.63);
\draw[draw=successgreen, line width=0.4pt] (11.60,-0.80) rectangle (12.64,-0.63);
\draw[draw=successgreen, line width=0.3pt] (12.12,-0.61) -- (12.12,-0.46);
\node[fglbl, text=successgreen, anchor=south] at (12.12,-0.46)
  {staging $\mathcal O(s\beta)$};
\node[fglbl, text=black!70, anchor=north] at (12.12,-1.32) {streamed};

\node[fglbl, text=black!70, anchor=north west, align=left] at (8.75,-1.56)
  {$12\Theta/W=116\gib$ at $\Theta=10^{12}$, $W=96$};

\end{tikzpicture}
  \caption{\streamadam. \textbf{(a)} Optimizer offload serializes one CPU update
  over all $\Theta/W$ parameters. \streamadam instead partitions $\Theta$ into buckets and rotating them through transfer, update, and write-back streams, leveraging communication-computation overlap. \textbf{(b)} Device bytes per rank: offload removes the $12\Theta/W$ optimizer state and
  streaming adds only $\mathcal O(s\beta)$ of staging.}
  \label{fig:main:streamadam-diagram}
\end{figure}
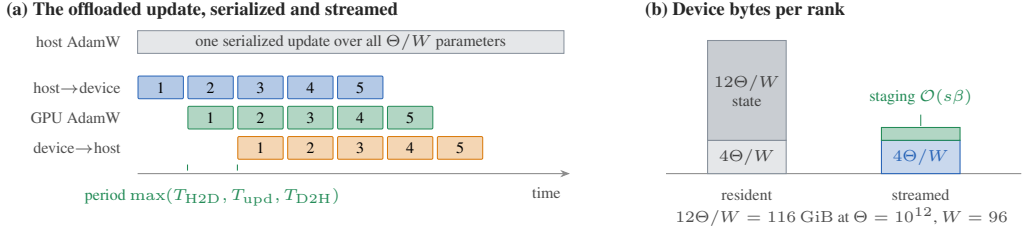

When HBM capacity is at the limit, it makes sense to offload the optimizer and its state to CPU \citep{ren2021zerooffload,zeroinfinity}, and the price is that a CPU AdamW update is slow, during that time the GPU device is idle and, having released its activations, largely empty in the HBM.  
We make use of the idle GPU as the update engine by streaming CPU-resident states to GPU and back.
Specifically, \streamadam partitions each
rank's state into buckets of at most $\beta$ parameters and rotates them
through $s$ staging slots and three streams: \Ni host-to-device transfer of
master weights, moments, and final gradients; \Nii a fused GPU AdamW update and
bf16 working-weight refresh; and \Niii write-back of updated fp32 state
(\cref{fig:main:streamadam-diagram}).  For $G$ buckets,
\begin{equation}
 T_{\mathrm{stream}} =
 G\max(T_{\mathrm{H2D}},T_{\mathrm{update}},T_{\mathrm{D2H}})
 +\mathcal O(T_{\mathrm{H2D}}+T_{\mathrm{update}}+T_{\mathrm{D2H}}),
 \quad M_{\mathrm{stage}}=\mathcal O(s\beta).
 \label{eq:main:adam}
\end{equation}
The floor for any offloaded optimizer is the host-link round trip of
$12\Theta/W$ bytes, so the design goal is to hide the update and the write-back
behind that transfer.  


\begin{table}[t]
\centering
\small
\caption{Optimizer step on gpt-oss-20b, 8xH200 GPUs. The baseline is the AVX-vectorized CPU Adam kernel of
ZeRO-Offload.  Both sides keep the same fp32 master weights and moments in host
memory.  Staging is resident memory per GPU.  Bucket-size sweeps are in
\cref{tab:streamadam-geometry-sweep}.
}
\label{tab:main:streamadam}
\setlength{\tabcolsep}{6pt}
\begin{tabular}{@{}lrrr@{}}
\toprule
Configuration & Step time (s) & Speedup & Staging (GiB/GPU) \\
\midrule
CPU Adam \citep{ren2021zerooffload} & 3.95 & --- & 0 \\
Streamed, 2 slots & \textbf{1.93} & $\mathbf{2.05\times}$ & \textbf{4.234} \\
\bottomrule
\end{tabular}
\end{table}

\paragraph{Isolated benchmark.}
\Cref{tab:main:streamadam} compares the \streamadam against the AVX CPU Adam kernel of
ZeRO-Offload \citep{ren2021zerooffload} that it replaces, with the same fp32
master weights and moments on the host in both cases. Rotating the same update
through the idle GPU takes the step from $3.95$ to $1.93$\,s, a
$2.05\times$ speedup.  Two slots are enough
to keep the transfer stream busy. Making the queue deeper but increasing staging slots $s$ leaves step time unchanged. That insensitivity to $s$ is consistent with the pipeline being transfer-bound at the $12\Theta/W$ host-link floor of \cref{eq:main:adam}.

\subsection{Composition and the Per-Rank Budget}
\label{sec:main:composition}

The four operators bound disjoint live sets: dispatch temporaries inside MoE
layers, logit strips at the output head, boundaries spanning layers, and
optimizer state at update time.  None requires any other, and each can be
enabled alone.  For the combined system we place them inside the \mop rank
layout \citep{nguyen2026mop}, which supplies the surrounding
component-specialized assignment of ZeRO-3 for dense weights, a
sequence-to-head all-to-all for attention, and expert-parallel placement for
expert weights.  The $W$ devices form overlapping sub-groups of sizes $D$,
$E_p$, and $P$ for sequence, expert, and vocabulary work, and every rank owns a
distinct token shard and a shard of each weight it touches, so the three
degrees are choices on one rank set rather than multiplicative axes of a device
equation $W=d\,t\,p$.  The per-rank device budget is then
\begin{equation}
M_{\mathrm{peak}}^{\mathrm{integrated}}=
  \underbrace{\tfrac{4\Theta}{W}}_{\text{weights, grads}}
  +\underbrace{\mathcal O(s\beta)}_{\text{opt.\ staging}}
  +\underbrace{\mathcal O(NHb)}_{\text{attention}}
  +\underbrace{\mathcal O(E_pqHb)}_{\text{dispatch}}
  +\underbrace{\mathcal O(NVb/P)}_{\text{strips}}
  +\underbrace{\mathcal O(NHb)}_{\text{boundaries}} .
\label{eq:main:total-peak}
\end{equation}
Every quantity on the right is fixed by model or launch configuration once
$N\le N_{\max}$, which is the property the four operators exist to deliver: a
configuration can be checked for feasibility before it is launched.  Two
distinct savings meet in the first two terms and are worth separating: host
offload supplies the reduction from $16\Theta/W$ to $4\Theta/W$ of persistent
state \citep{ren2021zerooffload}, while \streamadam reduces the time that
offload costs.

\section{End-to-End Integration}\label{sec:main:experiments}

\Cref{sec:main:method} priced each operator against the baseline that solves the same problem in isolation. This section runs all four together inside the \mop rank layout of \cref{sec:main:composition}, on three MoE models of 120B, 241B, and 667B parameters at 16, 32, and 64 H200 GPUs. The comparison is FSDP2-best, the highest-throughput configuration of a sweep over FSDP2 combined with expert, context, and tensor parallelism \citep{zhao2023fsdp}, on the same model at the same GPU count. Model dimensions are in \cref{app:scaling-configuration}.

\begin{figure}[t]
  \centering
  \includegraphics[width=\linewidth]{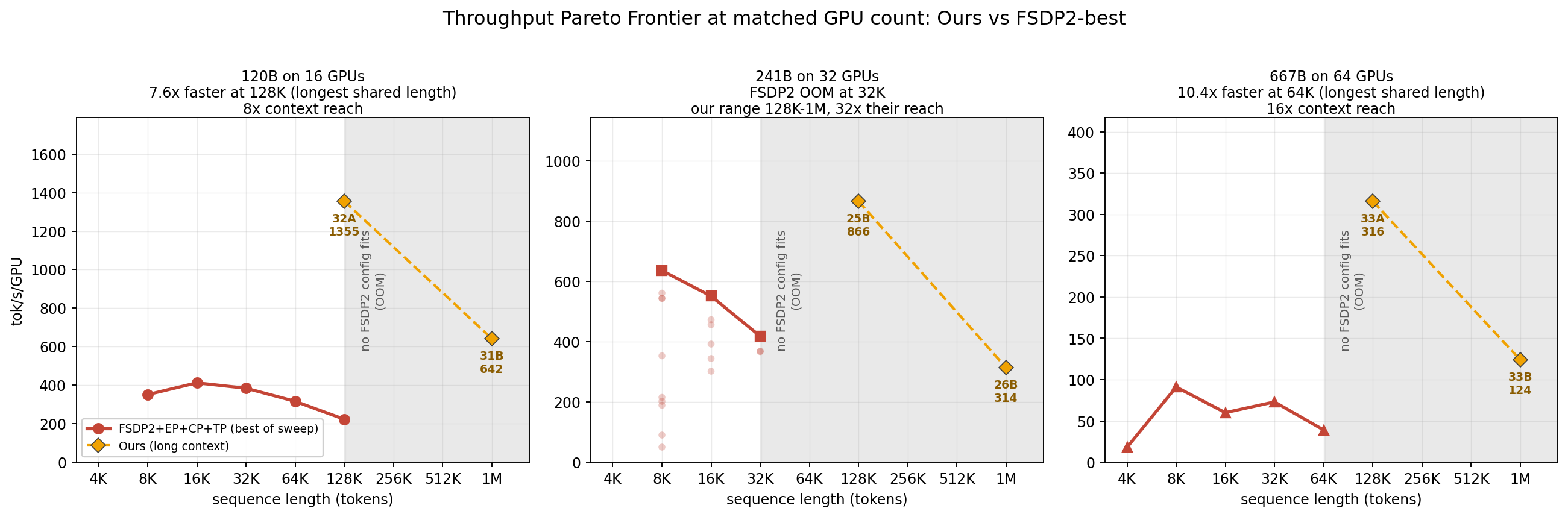}\\[3pt]
  \includegraphics[width=\linewidth]{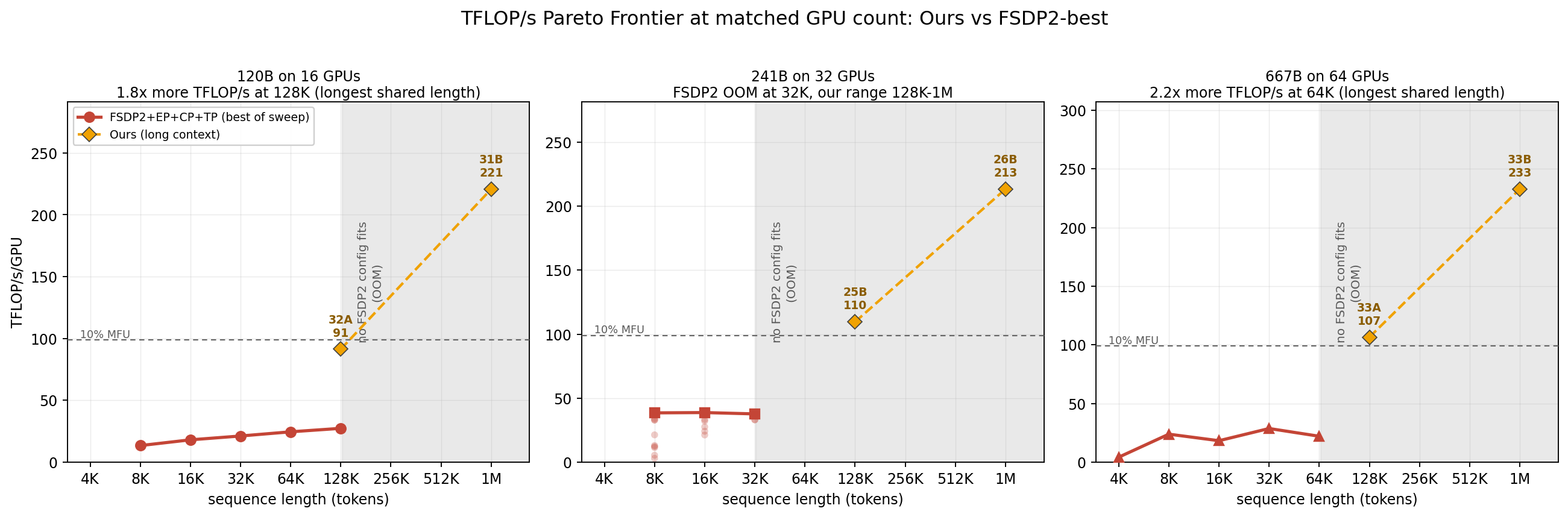}
  \caption{End-to-end training at matched model and GPU count. \textbf{Top:} per-GPU token throughput. \textbf{Bottom:} per-GPU floating-point rate, against a dashed line at $10\%$ model FLOP utilization. In both, the solid line is the best FSDP2 configuration at each length and the faint markers are the dominated ones from the same sweep, the diamonds are the composed stack, and the shaded band covers lengths at which no FSDP2 configuration fits. Token throughput falls with context while the floating-point rate rises, because attention work per token grows with context.}
  \label{fig:main:frontier}
\end{figure}

\paragraph{Context reach and throughput.}
\Cref{fig:main:frontier} (top) plots per-GPU token throughput against context length. FSDP2-best exhausts device memory past 128K, 32K, and 64K tokens at the three scales, while the composed stack trains at one million tokens at all three, which is $8$--$32\times$ the reach. It is also the faster of the two at the longest length FSDP2-best reaches, by $7.6\times$ at 128K on 120B and $10.4\times$ at 64K on 667B. At 241B, FSDP2-best stops at 32K, four times below our shortest configuration.
\Cref{fig:main:frontier} (bottom) reports TFLOPS/s. On that axis the trend reverses because of attention. Our per-GPU rate roughly doubles between 128K and 1M tokens at all three scales, from $91$--$110$ to $213$--$233$ TFLOP/s, while FSDP2-best stays below $40$ anywhere in its sweep. At those same lengths, the composed stack delivers $2.2\times$ the baseline's rate at 667B scale.

\begin{figure}[t]
  \centering
  \includegraphics[width=0.8\linewidth]{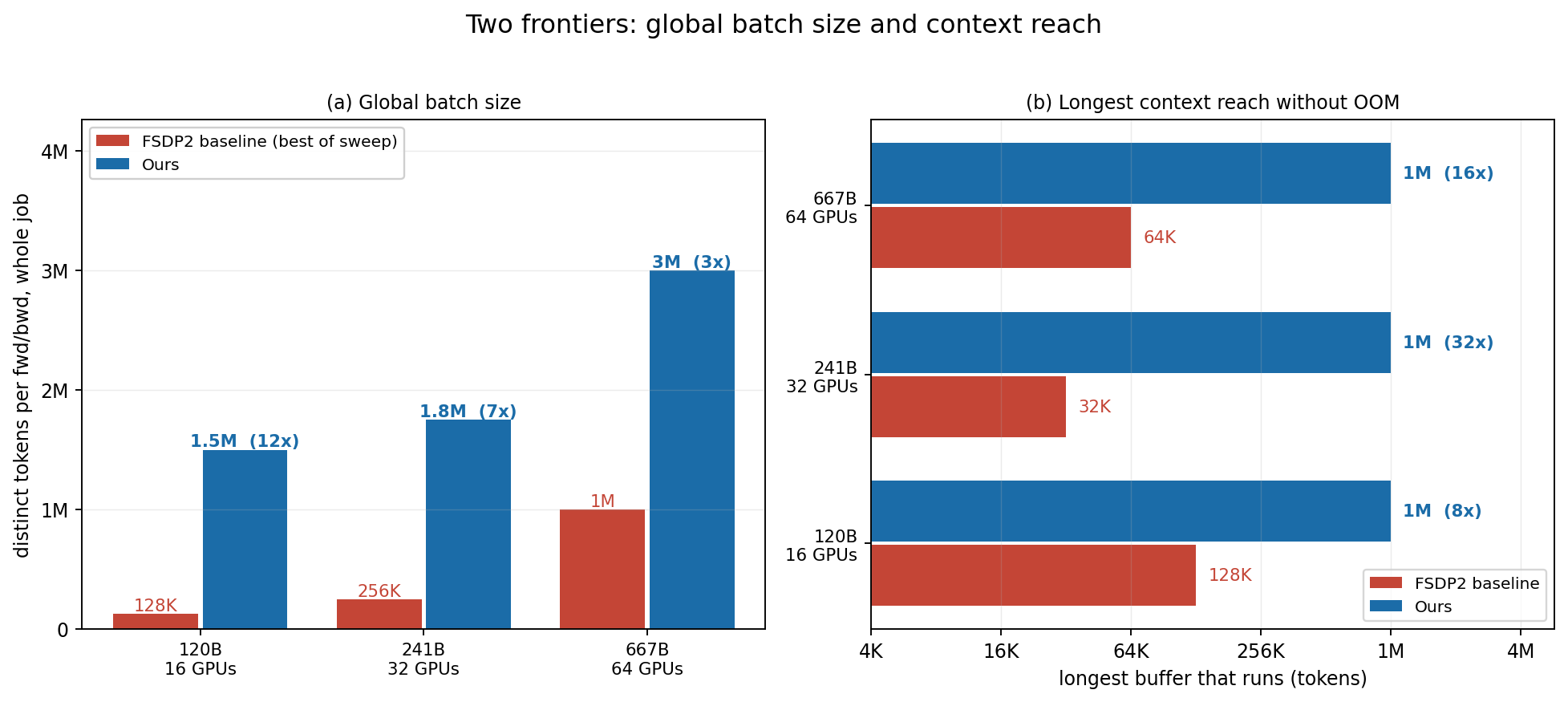}
  \caption{Batch and context frontiers at matched model and GPU count. \textbf{(a)} Largest global batch that runs, in distinct tokens per forward and backward pass. \textbf{(b)} Longest context that runs without exhausting device memory, on a log axis. Multiples are against FSDP2-best at the same scale.}
  \label{fig:main:frontier-batch}
\end{figure}

\paragraph{Batch size.}
Device memory that a step does not commit to a fixed working set is available for tokens, and \cref{fig:main:frontier-batch} gives the two axes it can be spent on. The largest global batch that runs is $1.5$M, $1.8$M, and $3$M distinct tokens per forward and backward pass, which is $12\times$, $7\times$, and $3\times$ the baseline's, and the longest context that runs is 1M tokens at every scale. Both follow from the same property. Every term of \cref{eq:main:total-peak} is fixed by model or launch configuration, so a workload that grows becomes more streamed work rather than a larger footprint. Training quality remains unchanged, as shown in \cref{app:quality}.

\section{Related Work}\label{sec:main:related}

\paragraph{Parallelism and expert dispatch.}
Megatron-style tensor, sequence, and pipeline parallelism factor devices into a
global grid \citep{shoeybi2019megatron,korthikanti2023reducing}, ZeRO and FSDP
shard model state \citep{rajbhandari2020zero,zhao2023fsdp}, expert parallelism
partitions sparse experts \citep{rajbhandari2022deepspeedmoe}, Ulysses shards
long sequences \citep{jacobs2023ulysses}, and \mop assigns a schedule per
component over overlapping sub-groups of one rank set \citep{nguyen2026mop};
this paper supplies the operator definitions, bounds, exactness conditions, and
mechanism-level evaluation that layout leaves unspecified.  FasterMoE, Lina,
PipeMoE, and ScheMoE partition dispatch to overlap communication, choosing the
partition count for bandwidth utilization
\citep{fastermoe,lina,pipemoe,schemoe}, which sizes the receive buffer before
the partition is applied and leaves the footprint following the realized load.
\lleptwo asks the different question of what a destination may receive, and
answers it with a token budget on each source, a chunk membership policy that a
relocated assignment plan requires, and a nested checkpoint contract that holds
the resulting bound through checkpointed backward.  Placement methods
redistribute skew across devices \citep{smartmoe,flexmoe,nguyen2026llep} and
compose with the budget, since a placement fitted to observed co-selection
statistics carries no guarantee for the next routing matrix.

\paragraph{Streamed losses, activations, and optimizer state.}
Cut Cross-Entropy evaluates the log-sum-exp on the fly in on-chip memory
\citep{wijmans2025cce} and Liger-Kernel fuses the projection with
cross-entropy over chunked inputs \citep{hsu2024liger}, both resting on the
online-softmax recurrence \citep{milakov2018online} behind FlashAttention
\citep{dao2022flashattention}, while Megatron's vocabulary-parallel
cross-entropy shards the weight and reduces per-shard statistics across
tensor-parallel ranks \citep{shoeybi2019megatron}.  Each assumes away the case
\ringtp addresses, since the fused kernels loop over a batch resident on one
device and the vocabulary-parallel reduction needs a replicated batch, so no
schedule is required; Ring Attention circulates key-value blocks around a ring
\citep{liu2024ringattention}, where no payload choice arises because only one
operand can move.  For activations, checkpointing and rematerialization reduce
retained interiors \citep{chen2016sublinear,checkmate,dtr}, hooks and swapping
relocate them \citep{pytorchautograd,capuchin}, and DeepSpeed
\citep{deepspeedcpucheckpointing} and PyTorch \citep{pytorchcheckpointwrapper}
offload checkpoint regions wholesale, whereas \selectoffload offloads a host-budget subset of checkpoint boundaries and prefetches them one layer ahead.  ZeRO-Offload and ZeRO-Infinity move
optimizer state and its update to lower tiers
\citep{ren2021zerooffload,zeroinfinity}, while 8-bit quantization
\citep{dettmers2022eightbit} and block-shared learning rates
\citep{zhang2025adammini} shrink it by changing numerics, which this setting
excludes.

\section{Conclusion}\label{sec:main:conclusion}

MoE training at long context or large batch size fails for four unrelated
reasons whose relative heights move with the configuration, so a stack is usable
only when every one of them has a bound and none of the bounds is mandatory.  We
gave each a schedule whose GPU working set follows from launch configuration: a
per-source token budget for dispatch, a ring of vocabulary meetings, a host-budget checkpoint-boundary offload, and a bucket pipeline for the offloaded
update.  Together they yield a closed-form per-rank budget in which every term is
fixed once the token ceiling is set, and matched component tests show each bound
holding at a throughput cost we measure.  Composed inside the \mop rank layout at
120B, 241B, and 667B, the stack trains at one-million-token context where a tuned
FSDP2 baseline runs out of memory between 32K and 128K, and spends the memory it
saves on a larger batch as readily as on a longer context.

\section{Limitations}\label{sec:main:limitations}

Four boundaries follow from occupying the memory-efficiency end of the space.
All-to-all and ring traffic assumes a fast interconnect, so the launch-count
trade that makes additional chunks inexpensive on the intra-node NVLink measured
here could differ across a slower fabric. A smaller token budget tightens the
receiver bound and raises the chunk count by the same factor, and forward time is
flat in the chunk count only over part of that range, so $c$ is selected from a
measured curve rather than minimized (\cref{fig:budget-selection}).  Checkpoint
and optimizer streaming spend host capacity and link bandwidth, which binds on
nodes with less host memory than the $2$\,TB used here. Topology
sensitivity and automatic selection of $(D,E_p,P,c,\beta)$ remain open.

\label{page:main-end}

\section*{The Use of Large Language Models}
Large language models were used as a writing and coding aid.  For the
manuscript, they assisted with drafting and editing prose, tightening captions,
and formatting references; every claim, number, and citation was checked
against the source measurements and the cited papers by the authors.  For the
implementation, they assisted with routine code editing and test scaffolding;
all operators, correctness tests, and measurements were designed, reviewed, and
executed by the authors.  Large language models were not used to generate
research ideas, experimental designs, or results, and the authors take full
responsibility for the content of this paper.

\bibliography{references}
\bibliographystyle{plainnat}

\newpage
\appendix
\section{Experimental Setup}\label{app:setup}

Each operator is measured on the live set it governs, so each benchmark is configured for that operator rather than for a shared end-to-end workload; \cref{app:full-configuration} collects the settings. Every controlled benchmark runs on one node with 8 NVIDIA H200 GPUs and compares systems on byte-identical input in the same execution, so a comparison never spans two allocations of the machine. Latencies for the dispatch benchmark are means over ten measured forward calls after three warmup calls. Memory is the maximum PyTorch peak allocation across ranks for the dispatch and projection benchmarks, and the maximum 100\,ms NVML sample across ranks after warmup for \selectoffload, since no single counter captures allocator peaks, external runtime allocations, and host residency at once. Tensor totals count each allocation once.

\begin{table}[ht]
\centering
\small
\caption{Configuration of each controlled component benchmark.}
\label{app:full-configuration}
\begin{tabular}{L{0.27\textwidth} L{0.65\textwidth}}
\toprule
Benchmark & Configuration \\
\midrule
Hardware & One node, eight NVIDIA H200 GPUs, intra-node NVLink, 2\,TB host memory \\
Expert dispatch (\lleptwo) & Two shapes: 32{,}768 tokens per rank at $H=I=4096$ with top-$4$ and $c=10{,}923$, and 65{,}536 tokens per rank at $H=7168$, $I=2048$ with top-$8$ and $c=6554$; 128 experts, one eight-rank expert-parallel group, BF16; strided chunk membership; expert-matmul checkpointing; $384\mib$ local-expert workspace \\
Vocabulary projection (\ringtp) & A distinct local batch of $N=16{,}384$ per rank, $H=7168$, $V=200{,}000$, FP32; projection group $P\in\{4,8\}$; auto-mode operand selection \\
Checkpoint offload (\selectoffload) & gpt-oss-20b in BF16; sequence and expert parallelism of degree eight; 24 decoder layers exposing 47 checkpoint boundaries; $556{,}432$ tokens per step at a configured ceiling of $557{,}056$; pinned host budget swept over $\{0,8,16\}\gib$ and unlimited; depth-one restore prefetch \\
Offloaded optimizer (\streamadam) & gpt-oss-20b; fp32 master weights and moments in host memory on both sides; 100M-parameter buckets and $s=2$ staging slots, selected from \cref{tab:streamadam-geometry-sweep} \\
\bottomrule
\end{tabular}
\end{table}

\paragraph{Scaling workload.}\label{app:scaling-configuration}
The end-to-end comparison of \cref{sec:main:experiments} is a pre-training workload, with all weights initialized from scratch, at 120B, 241B, and 667B parameters on 2, 4, and 8 nodes of the same type, that is 16, 32, and 64 GPUs. The three models share one architecture and differ only in depth: hidden width $H=7168$, dense FFN intermediate width $18{,}432$, 384 routed experts of intermediate width $I=2048$ with top-$8$ routing, and $V=200{,}000$. The 120B, 241B, and 667B models have 7, 14, and 39 decoder layers. Routing is dropless, so every token reaches all eight of its experts, and load is balanced by the auxiliary-loss-free rule of DeepSeek-V3, which adds a per-expert bias to the routing scores and adjusts each bias at the end of every step according to that expert's load \citep{deepseekv3}. In the layout of \cref{app:mop}, expert shards span every rank, so $E_p=W$ is 16, 32, and 64 at the three scales, the projection group is $P=8$, and $D$ is set between 2 and 16 according to the context length. Both stacks run the same model at the same GPU count at each scale. FSDP2-best is selected per scale by sweeping FSDP2, tensor, context, and expert parallelism together with optimizer CPU offload on a TorchTitan-based stack \citep{zhao2023fsdp,liang2025torchtitan}, and taking the highest-throughput configuration that fits; the whole sweep appears in \cref{fig:main:frontier} as faint markers, and its per-length upper envelope is the solid line.
The per-source token budget $c$ is sweeped from 4096 to 32768 tokens per rank.

\paragraph{Rank layout.}\label{app:mop}
\mop assigns a parallelism scheme per model component instead of factoring the devices into a grid \citep{nguyen2026mop}. The same $W$ ranks form three overlapping sub-groups: $D$ ranks share a sequence and trade it for attention heads through an all-to-all, $E_p$ ranks hold disjoint expert shards, and $P$ ranks hold disjoint vocabulary intervals, with dense weights sharded by ZeRO-3 across all $W$ \citep{rajbhandari2020zero}. Because the sub-groups are views of one rank set rather than factors of a device equation $W=d\,t\,p$, the three degrees are chosen independently of one another, and each operator of \cref{sec:main:method} runs inside the sub-group that owns the live set it bounds.

\paragraph{Training quality.}\label{app:quality}
Each operator preserves forward values and gradients, so the composed stack trains to the same quality as a standard implementation given the same recipe. We check this end to end by supervised fine-tuning gpt-oss-20b at low reasoning effort on Nemotron-Math \citep{nemotronmath}, whose assistant responses are generated by gpt-oss-120b, at a context length of 128K, and evaluating Avg@8 on AIME 2025. Trained on the composed stack the model reaches $59.8\%$; trained on the FSDP2 baseline under the same data, schedule, and hyperparameters it reaches $59.6\%$; and gpt-oss-20b before fine-tuning reaches $45.2\%$.

\section{\lleptwo}

\subsection{Backward and the Nested Checkpoint Contract}\label{app:llep-backward}

The peak of a checkpointed step falls inside backward, so \cref{eq:main:llep-bound} is only useful if it survives recomputation. Three properties deliver that.

\paragraph{The partition is recomputable, not saved.}
Chunk $i$ holds token positions $\{i,i+K,i+2K,\dots\}\cap[0,N)$, a function of $N$ and $K$ alone, and $K$ is reduced with a maximum over the expert-parallel group before any chunk collective is issued. A forward and its recomputation therefore cut the token axis identically without storing anything route-scale, and every rank runs the same number of chunks even when their token counts differ.

\paragraph{The inner checkpoint is reentrant so the outer one can see it.}
The expert matmul holds the $\Theta(k\ceff I)$ expert intermediate and the stacked expert weights the chunk needs, and it is wrapped in a \emph{reentrant} checkpoint. Reentrant checkpointing routes every tensor argument through autograd's saved-tensor mechanism \citep{pytorchautograd}, so when the decoder layer's own non-reentrant checkpoint is active its pack hook sees each of those tensors and replaces it with a placeholder. The stacked weights are then resident for one layer's forward rather than from that forward until the global backward. A non-reentrant inner checkpoint would instead keep the forward function and its arguments in an unpack closure holding strong Python references for the whole forward-to-backward lifetime; those references sit outside any saved-tensor mechanism, so no outer hook can intercept them, and with $K$ chunks over $L$ layers the stacked weights would stay live for $\mathcal O(KL)$ cross-layer duration. Only the expert matmul is checkpointed, not the whole chunk, because placing all three stages of chunk $i$ in one checkpoint call serializes them from Python's point of view and prevents the dispatch prefetch of chunk $i{+}1$; the cost is that dispatch outputs stay live across chunks, which \cref{eq:main:llep-bound} already bounds.

\paragraph{Gradients accumulate once per layer.}
Dispatch and combine use all-to-all primitives whose backward is the symmetric inverse all-to-all, so gradients flow through the pipeline as ordinary autograd tensors and no chunk graph is retained past its own backward. Every chunk's expert-matmul backward accumulates into the same stacked expert-weight tensor, so the weight-migration backward fires exactly once per layer and expert-parameter gradients are returned once, preserving a single sharded-optimizer hook boundary. A scalar anchor returned by the weight-transfer function is added to the final output once rather than per chunk, which gives a rank that only sends foreign expert weights and never receives routes a path from the loss to its transfer.

\begin{algorithm}[t]
\caption{\lleptwo on one expert-parallel rank}
\label{alg:llep-v2}
\begin{algorithmic}[1]
  \State {\bfseries Input:} hidden states $X$, route weights $A$, expert ids $E$, local
  expert parameters $\Theta_{\mathrm{loc}}$, token budget $c$, ceiling $K_{\max}$
  \State $K\leftarrow\min(\mathrm{AllReduceMax}(\lceil N/c\rceil),K_{\max})$
  \State run the LLEP load measurement and assignment plan once; migrate expert weights if
  the plan requires it
  \State $\pi\leftarrow$ per-route destinations from the layer-wide plan;
  $S\leftarrow\mathrm{AllGather}$ of the $(K,E_p)$ send splits
  \State $Y\leftarrow 0$; issue dispatch of chunk $0$
  \For{$i=0,\ldots,K-1$}
    \State $\mathcal{T}_i\leftarrow\{i,i+K,i+2K,\ldots\}\cap[0,N)$
    \Comment{strided membership}
    \State $(X_i,A_i,E_i)\leftarrow$ gather rows $\mathcal{T}_i$ in destination order
    under $\pi$
    \State $(\widehat X_i,\widehat A_i,\widehat E_i)
    \leftarrow\mathrm{A2A}((X_i,A_i,E_i);S_i,S_i^{\top})$
    \Comment{dispatch, one coalesced group}
    \State $\widehat Y_i\leftarrow\mathrm{ckpt}_{\mathrm{reentrant}}
    \left[\mathrm{LocalExperts}(\widehat X_i,\widehat A_i,\widehat E_i;
    \Theta_{\mathrm{loc}})\right]$
    \State issue dispatch of chunk $i{+}1$ on the dispatch stream
    \State $Y_i\leftarrow\mathrm{A2A}(\widehat Y_i;S_i^{\top},S_i)$
    \Comment{combine}
    \State $Y[\mathcal{T}_i]\leftarrow$ top-$k$ reduction of $Y_i$; release chunk-local
    tensors
  \EndFor
  \State {\bfseries Output:} $Y$, and the anchor that drives the weight-transfer backward
  once
\end{algorithmic}
\end{algorithm}

\paragraph{Which gradient the partition touches.}
For $g_t=\partial\mathcal L/\partial y_t$, route weight $\alpha_{tj}$, and expert function $f_e$,
\begin{equation}
\frac{\partial\mathcal L}{\partial\alpha_{tj}}=g_t^\top f_{e(t,j)}(x_t),\qquad
\frac{\partial\mathcal L}{\partial x_t}=\sum_j\alpha_{tj}J_{f_{e(t,j)}}(x_t)^\top g_t,\qquad
\frac{\partial\mathcal L}{\partial\theta_e}=\!\!\sum_{(t,j):e(t,j)=e}\!\!
  \alpha_{tj}\frac{\partial f_e(x_t)}{\partial\theta_e}^{\!\top}\! g_t .
\label{eq:route-gradients}
\end{equation}
Only $\partial\mathcal L/\partial\theta_e$ has its summation order changed, since it is the one sum split across chunks and reassembled in the shared accumulator of \cref{alg:llep-v2}. At $K=1$ that sum has a single term and the expert-parameter gradients are bitwise identical to LLEP's. Hidden-state and router gradients are never split.

\paragraph{Cost accounting.}
Route payload is unchanged in both directions: each route crosses the interconnect once per pass, and the additions are per-chunk metadata exchanges that are $\mathcal O(1)$ per route rather than $\mathcal O(H)$. Two costs scale with $K$. Collective launches grow linearly, since dispatch issues one coalesced group of three native-dtype all-to-alls, combine issues one, and backward issues their symmetric inverses. The expert matmul is evaluated three times per update under decoder-level checkpointing, namely the original forward, the decoder recomputation, and the inner recomputation, against twice for a checkpointed unpartitioned layer; dispatch and combine are evaluated twice either way.

\subsection{Routing Profiles}\label{app:llep-profiles}

A profile is a pair $r/h$ in which $r$ is the hot-token percentage and $h$ the number of hot experts. Experts $0,\dots,h-1$ are hot and the remaining $E-h$ are cold. The first $\lfloor rN\rfloor$ token positions are hot and select experts $\{j\bmod h: j=0,\dots,k-1\}$, so a hot token's $k$ routes cycle over the hot set and collapse onto $\min(k,h)$ distinct experts. Each remaining token $i$ selects cold experts $\{h+((i+j)\bmod(E-h)): j=0,\dots,k-1\}$, which walks the cold set and spreads the residual load uniformly. The balanced profile assigns token $i$ the experts $\{(i+j)\bmod E: j=0,\dots,k-1\}$. Routing weights are $1/k$ on every selected expert. Skew rises as $r$ rises and as $h$ falls, and $h\ge k$ against $h<k$ is the qualitative boundary, since below it the hot set cannot absorb $k$ distinct routes and repeats. We sweep $r/h\in\{30/16,50/16,80/16,95/16,30/4,50/4\}$ plus balanced.

\subsection{Per-Profile Measurements}

Unless a table states otherwise, every \lleptwo entry uses strided chunk membership and expert-matmul checkpointing, and the chunk count is stated per table. Peak memory is deterministic in the configuration at both shapes: three repeats of the 65K sweep reproduce all seven peaks to the milli-GiB, and latency spread across those repeats is at most $2.6\%$.

\begin{table}[ht]
\centering
\small
\setlength{\tabcolsep}{2pt}
\caption{Routing-skew sweep at both shapes.  Entries are mean forward latency in ms /
maximum peak allocated memory across ranks in GiB, from executions in which the methods
compared run on identical input.  Speedup and peak reduction are against LLEP.  At 65K the
sweep was repeated three times and each row reports the repeat with the lowest \lleptwo
latency together with that repeat's own LLEP measurement, so the columns are paired;
\cref{app:llep-latency-modes} explains why the lowest repeat is representative.  At the 65K
shape standard expert parallelism runs out of memory on every skewed profile; on the
balanced profile LLEP performs no relocation, so the two coincide.}
\label{tab:llep-sweeps-full}
\begin{tabular}{@{}L{0.11\textwidth} R{0.155\textwidth} R{0.155\textwidth} R{0.155\textwidth} R{0.115\textwidth} R{0.125\textwidth}@{}}
\toprule
Routing & Standard EP (ms / GiB) & LLEP (ms / GiB) & \lleptwo (ms / GiB) & Speed vs.\ LLEP & Peak saved \\
\midrule
\multicolumn{6}{@{}l}{\emph{32{,}768 tokens per rank, $H=I=4096$, top-$4$, $c=10{,}923$ ($K=3$)}}\\
Balanced & 40.82 / 21.347 & 47.95 / 24.101 & 44.74 / 17.606 & $\mathbf{1.07\times}$ & \textbf{26.9\%} \\
30\% / 16 & 89.39 / 32.549 & 53.14 / 24.476 & 51.44 / 18.460 & $1.03\times$ & 24.6\% \\
50\% / 16 & 143.12 / 45.351 & 53.74 / 24.476 & 54.57 / 18.460 & $0.98\times$ & 24.6\% \\
80\% / 16 & 225.06 / 64.556 & 53.89 / 24.664 & 53.90 / 18.553 & $1.00\times$ & 24.8\% \\
95\% / 16 & 267.08 / 74.154 & 54.34 / 24.664 & 53.93 / 18.553 & $1.01\times$ & 24.8\% \\
30\% / 4 & 107.28 / 36.885 & 53.49 / 24.664 & 52.27 / 18.554 & $1.02\times$ & 24.8\% \\
50\% / 4 & 157.39 / 48.445 & 53.61 / 24.664 & 53.97 / 18.554 & $0.99\times$ & 24.8\% \\
\midrule
\multicolumn{6}{@{}l}{\emph{65{,}536 tokens per rank, $H=7168$, $I=2048$, top-$8$, $c=6554$ ($K=10$)}}\\
Balanced & 177.82 / 52.660 & 177.82 / 52.660 & 161.00 / 21.430 & $\mathbf{1.10\times}$ & \textbf{59.3\%} \\
30\% / 16 & OOM & 182.97 / 52.906 & 176.62 / 22.772 & $1.04\times$ & 57.0\% \\
50\% / 16 & OOM & 186.07 / 52.906 & 182.64 / 22.773 & $1.02\times$ & 57.0\% \\
80\% / 16 & OOM & 189.14 / 52.988 & 180.30 / 22.861 & $1.05\times$ & 56.9\% \\
95\% / 16 & OOM & 182.13 / 52.988 & 178.57 / 22.855 & $1.02\times$ & 56.9\% \\
30\% / 4 & OOM & 185.78 / 52.906 & 180.59 / 22.774 & $1.03\times$ & 57.0\% \\
50\% / 4 & OOM & 186.79 / 52.906 & 184.69 / 22.778 & $1.01\times$ & 57.0\% \\
\bottomrule
\end{tabular}
\end{table}

\begin{table}[ht]
\centering
\small
\setlength{\tabcolsep}{3pt}
\caption{Micro-batch length ladder at the 65K shape under one token budget, $c=4096$ with
$K_{\max}=10$, against a chunk count fixed at $K=10$.  Speedup and peak reduction compare
\lleptwo with LLEP at the same length, with LLEP re-measured in the same execution; ranges
are over the balanced and $80\%/16$ profiles.  Where a length was executed more than once
we report the lowest speedup observed, so the ranges are the pessimistic reading.  The
ceiling binds above $N=cK_{\max}=40{,}960$, where the two configurations coincide.}
\label{tab:llep-length}
\begin{tabular}{@{}R{0.13\textwidth} R{0.09\textwidth} R{0.17\textwidth} R{0.15\textwidth} R{0.17\textwidth} R{0.15\textwidth}@{}}
\toprule
& \multicolumn{3}{c}{Token budget $c=4096$} & \multicolumn{2}{c}{Fixed $K=10$} \\
\cmidrule(lr){2-4}\cmidrule(lr){5-6}
Tokens $N$ & $K$ & Speedup & Peak saved & Speedup & Peak saved \\
\midrule
128     & 1  & $\mathbf{1.60}$--$\mathbf{2.96\times}$ & 9.0--11.0\%   & $0.45$--$0.48\times$ & 9.4--11.4\% \\
1,024   & 1  & $\mathbf{1.35}$--$\mathbf{2.17\times}$ & 10.4--11.1\%  & $0.47$--$0.56\times$ & 14.0--15.6\% \\
8,192   & 2  & $0.98$--$1.09\times$ & 14.3--17.0\%  & $0.80$--$0.82\times$ & 28.6\% \\
32,768  & 8  & $1.02$--$1.09\times$ & 44.9--47.6\%  & $0.98$--$1.06\times$ & 47.2--49.4\% \\
65,536  & 10 & $1.04$--$1.10\times$ & 56.9--59.3\%  & \multicolumn{2}{c}{\textit{coincides}} \\
131,072 & 10 & $1.07$--$1.10\times$ & 63.4--65.9\%  & \multicolumn{2}{c}{\textit{coincides}} \\
\bottomrule
\end{tabular}
\end{table}

A chunk count tuned for the top of the range cuts a 128-token micro-batch into ten pieces and pays ten full collective rounds for roughly one token each, which costs $0.45$--$0.48\times$; the budget derives $K=1$ instead and runs at $1.60$--$2.96\times$, faster than LLEP itself, because the single chunk still uses the coalesced single-group all-to-all. Peak reduction tracks the size of the unpartitioned live set, rising from $9.0\%$ at 128 tokens to $65.9\%$ at 131{,}072, since below roughly 8{,}192 tokens per rank the dispatch buffers are small next to the resident expert weights.

\begin{table}[ht]
\centering
\small
\setlength{\tabcolsep}{4pt}
\caption{Chunk membership at 65{,}536 tokens per rank, $K=10$, all other settings
identical.  \emph{Send ratio} is the largest destination send divided by the mean send
within a chunk, taken as the maximum over chunks and measured by a per-chunk counter on the
dispatch all-to-all.  Membership changes neither the token set nor any route's destination,
so the two policies differ only in schedule quality.}
\label{tab:llep-membership}
\begin{tabular}{@{}L{0.14\textwidth} R{0.13\textwidth} R{0.13\textwidth} R{0.13\textwidth} R{0.13\textwidth} R{0.10\textwidth} R{0.10\textwidth}@{}}
\toprule
& \multicolumn{2}{c}{Latency (ms)} & \multicolumn{2}{c}{Peak (GiB)}
& \multicolumn{2}{c}{Send ratio} \\
\cmidrule(lr){2-3}\cmidrule(lr){4-5}\cmidrule(lr){6-7}
Routing & Strided & Contig. & Strided & Contig. & Strided & Contig. \\
\midrule
Balanced  & \textbf{161.2} & 218.2 & \textbf{21.430} & 21.464 & 1.00 & 1.02 \\
95\% / 16 & \textbf{178.6} & 184.9 & \textbf{22.855} & 23.575 & 1.35 & 2.63 \\
80\% / 16 & \textbf{180.6} & 209.7 & \textbf{22.861} & 23.514 & 2.40 & 4.00 \\
50\% / 4  & \textbf{190.8} & 217.0 & \textbf{22.778} & 24.151 & 2.29 & 6.00 \\
\bottomrule
\end{tabular}
\end{table}

Strided membership runs at $1.03$--$1.35\times$ the speed of contiguous membership with $0.03$--$1.37\gib$ lower peak, and the send-ratio columns give the mechanism. The same counter reports $\max_dR_d^{(i)}$ against the bound of \cref{eq:main:llep-bound}. Strided membership realizes $12.5\%$ of it, which is the even share $1/E_p$, at every chunk size and on every profile, while contiguous membership reaches $17.1\%$ on the $50\%/4$ profile. The bound is therefore respected everywhere and attained nowhere: all-to-all buffers are sized from realized split counts, so the peak follows $k\ceff$ and the bound is the guarantee that no router can move it rather than a description of the traffic.

\begin{table}[ht]
\centering
\small
\setlength{\tabcolsep}{5pt}
\caption{Forward and backward together, $c$ chosen to give $K=10$ at each shape, routing
profiles balanced / $80\%/16$.  Peak is the maximum allocated memory across ranks over the
combined pass.}
\label{tab:llep-fwd-bwd}
\begin{tabular}{@{}L{0.26\textwidth} R{0.16\textwidth} R{0.16\textwidth} R{0.13\textwidth} R{0.13\textwidth}@{}}
\toprule
Shape & LLEP (GiB) & \lleptwo (GiB) & Peak saved & Speedup \\
\midrule
65K tok, $H{=}7168$, top-8 & 52.660 / 52.988 & \textbf{21.430 / 22.861} & \textbf{59.3 / 56.9\%} & $1.11$ / $1.04\times$ \\
32K tok, $H{=}4096$, top-4 & 23.850 / 24.413 & \textbf{14.902 / 15.382} & \textbf{37.5 / 37.0\%} & $0.87$ / $0.90\times$ \\
\bottomrule
\end{tabular}
\end{table}

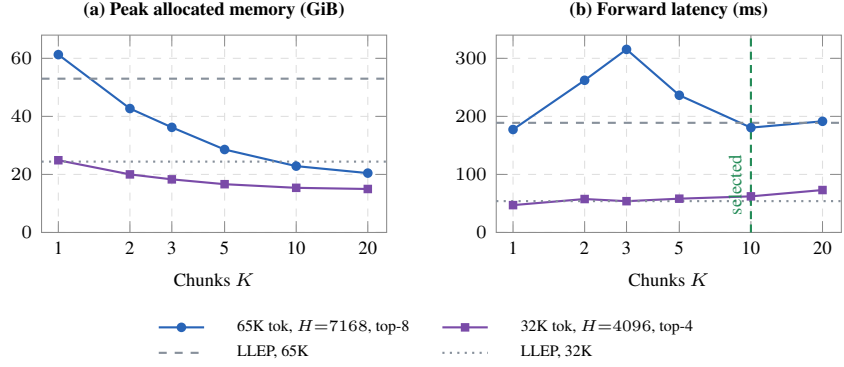
\begin{figure}[ht]
\centering
\begin{tikzpicture}
\begin{groupplot}[
  group style={group size=2 by 1, horizontal sep=1.45cm},
  width=0.44\textwidth, height=0.30\textwidth,
  xmode=log, log basis x=2,
  xtick={1,2,3,5,10,20}, xticklabels={1,2,3,5,10,20},
  xmin=0.85, xmax=24,
  xlabel={Chunks $K$},
  tick label style={font=\scriptsize},
  label style={font=\scriptsize},
  title style={font=\scriptsize\bfseries, yshift=-1.2mm},
  grid=major, grid style={dashed,gray!25},
  axis line style={black!45}, tick style={black!45},
  legend style={font=\fontsize{6}{7}\selectfont, draw=none, fill=none,
                at={(1.12,-0.40)}, anchor=north, legend columns=2,
                column sep=10pt, inner sep=1pt},
  legend cell align=left
]
\nextgroupplot[title={(a) Peak allocated memory (GiB)},
  ymin=0, ymax=68, ytick={0,20,40,60}]
\addplot[routeblue, mark=*, mark size=1.4pt, thick] coordinates {
  (1,61.273) (2,42.701) (3,36.222) (5,28.585) (10,22.861) (20,20.454)};
\addlegendentry{65K tok, $H{=}7168$, top-8}
\addplot[vocabpurple, mark=square*, mark size=1.2pt, thick] coordinates {
  (1,24.887) (2,20.010) (3,18.302) (5,16.634) (10,15.382) (20,14.983)};
\addlegendentry{32K tok, $H{=}4096$, top-4}
\addplot[baselinegray, dashed, thick] coordinates {(0.85,52.988) (24,52.988)};
\addlegendentry{LLEP, 65K}
\addplot[baselinegray, dotted, thick] coordinates {(0.85,24.413) (24,24.413)};
\addlegendentry{LLEP, 32K}

\nextgroupplot[title={(b) Forward latency (ms)},
  ymin=0, ymax=340, ytick={0,100,200,300}]
\addplot[routeblue, mark=*, mark size=1.4pt, thick] coordinates {
  (1,177.4) (2,262.2) (3,315.4) (5,236.4) (10,180.6) (20,191.5)};
\addplot[vocabpurple, mark=square*, mark size=1.2pt, thick] coordinates {
  (1,47.1) (2,57.6) (3,53.9) (5,58.2) (10,62.2) (20,73.1)};
\addplot[baselinegray, dashed, thick] coordinates {(0.85,188.9) (24,188.9)};
\addplot[baselinegray, dotted, thick] coordinates {(0.85,54.0) (24,54.0)};
\draw[successgreen, thick, dash pattern=on 3pt off 2pt]
  (axis cs:10,0) -- (axis cs:10,340);
\node[font=\scriptsize, text=successgreen, anchor=south east, rotate=90]
  at (axis cs:10,150) {selected};
\end{groupplot}
\end{tikzpicture}
\caption{Chunk count against peak memory and forward latency at both shapes, routing
profile $80\%/16$, with LLEP at the same shape as the horizontal reference.  Peak falls
monotonically in $K$ with decaying marginal return: at 65K the first ten chunks remove
$38.4\gib$ and the next ten remove $2.4\gib$.  Latency does not follow the same shape, so
$K$ is selected from this curve rather than minimized.  At 65K the points
$K\in\{2,3,5\}$ are slower and larger than $K=10$, while $K\in[10,20]$ is flat to within
$6\%$; at 32K, latency rises steadily past $K=3$.  $K=1$ separates the schedule from the
chunking: with one chunk the operator holds $8.3\gib$ more than LLEP at 65K, because a
single chunk adds a copy of the route-shaped tensors and partitions nothing.  Each point is
one execution of ten timed forward calls; LLEP is re-measured inside every execution and
its reference line is the mean of those six measurements, which span $0.3\%$.}
\label{fig:budget-selection}
\end{figure}

\Cref{tab:llep-fwd-bwd} confirms that the reduction is not an artifact of where a forward-only measurement stops: including backward, the reduction at the reference shape is unchanged at $56.9$--$59.3\%$. The 32K row shows the other side of the trade. At $K=10$ that shape sits past the flat region of \cref{fig:budget-selection}b, so the chunk count that is free at 65K costs $0.87$--$0.90\times$ there, which is what deriving $K=8$ from the budget avoids at that length.

\subsection{Choosing the Token Budget}\label{app:llep-budget}

\subsection{Two Latency Modes at the 65K Shape}\label{app:llep-latency-modes}

Forward latency at the 65K shape is bimodal, and the modes are far enough apart to matter for how the timing tables should be read. Five back-to-back passes over the seven profiles at $K=16$ separate them cleanly: in the fast mode every profile reproduces to within $0.5\%$ across passes, which is tighter than any other latency we measure, while the slow mode adds $12$--$56\%$ and never subtracts. The mode is not a property of the routing profile, since the same profile appears in either mode across passes, and it is strongly a property of the chunk count. Over everything measured at this shape, $K=10$ is slow in $2$ of $56$ profile measurements against $21$ of $84$ for $K=16$.

The mechanism is the host-side launch pattern. Enabling a per-chunk diagnostic counter, which inserts one host-side synchronization per chunk and changes nothing about the computation, moves the slow mode off two profiles and onto a third, reproducibly in both passes of both conditions. For a given chunk count, routing profile, and host synchronization structure the operator enters one mode and stays in it. Two consequences follow. The representative latency for a configuration is the fast mode, so \cref{tab:llep-sweeps-full} reports the lowest of three repeats at 65K. And $K_{\max}=10$ keeps the schedule in the region where the slow mode is rare, which is a second reason for the ceiling alongside the flat-region argument of \cref{fig:budget-selection}b. The single-execution points of \cref{fig:budget-selection}b at $K\in\{2,3,5\}$ are exposed to the effect: the hump at $K=3$ is $1.75\times$ the $K=10$ point, which exceeds the slow mode's range, so the ordering of those points against $K=10$ holds while their magnitudes are not tight.

\section{\ringtp}\label{app:ringtp}

Each rank holds a distinct local batch, and the meetings visit every batch with every vocabulary shard exactly once. Target-column ownership is exclusive: exactly one shard interval contains a given target index, so exactly one meeting contributes $y_t$ and no reduction over duplicated contributions is needed. Backward replays the same meetings from the saved per-token state, forms the softmax residual for one strip at a time, and contracts it into input and weight gradients, so neither pass stores a full logit or logit-gradient tensor.

\begin{algorithm}[ht]
\caption{\ringtp forward on rank $r$ of a projection group of size $P$}
\label{alg:ringtp}
\begin{algorithmic}[1]
  \State {\bfseries Input:} local batch $X_r$ of $N$ tokens, targets $t_r$, weight shard
  $W_r$ over the vocabulary interval $\mathcal V_r$
  \Function{Fold}{state $S$, strip $Y$, targets $t$, interval $\mathcal V$}
    \For{each token $n$}
      \State $(m,z,y)\leftarrow S_n$
      \State $m'\leftarrow\max\!\big(m,\max_v Y_{n,v}\big)$, \qquad
             $z'\leftarrow e^{m-m'}z+\sum_v e^{Y_{n,v}-m'}$
      \State \textbf{if} $t_n\in\mathcal V$ \textbf{then} $y\leftarrow Y_{n,t_n}$
      \State $S_n\leftarrow(m',z',y)$
    \EndFor
    \State \Return $S$
  \EndFunction
  \State $S\leftarrow\Call{Fold}{(-\infty,0,0),\;X_rW_r,\;t_r,\;\mathcal V_r}$
  \If{$N>V/P$}
    \State $W\leftarrow W_r$
    \For{$i=1,\ldots,P-1$}
      \State $q\leftarrow(r-i)\bmod P$
      \State send $W$ to rank $r{+}1$, receive $W_q$ from rank $r{-}1$, and set
      $W\leftarrow W_q$
      \State $S\leftarrow\Call{Fold}{S,\;X_rW,\;t_r,\;\mathcal V_q}$, then release the
      strip
    \EndFor
  \Else
    \State $(X,t,\widetilde S)\leftarrow(X_r,t_r,S)$
    \For{$i=1,\ldots,P-1$}
      \State send $(X,t,\widetilde S)$ to rank $r{+}1$ and receive the next
      $(X,t,\widetilde S)$ from rank $r{-}1$
      \State $\widetilde S\leftarrow\Call{Fold}{\widetilde S,\;XW_r,\;t,\;\mathcal V_r}$,
      then release the strip
    \EndFor
    \State send $\widetilde S$ to rank $r{+}1$ and receive $S$ from rank $r{-}1$
  \EndIf
  \State {\bfseries Output:} the negative log-likelihood $m+\log z-y$ of every token, from
  its state $S_n=(m,z,y)$
\end{algorithmic}
\end{algorithm}

\begin{figure}[ht]
  \centering
  \includegraphics[width=0.86\linewidth]{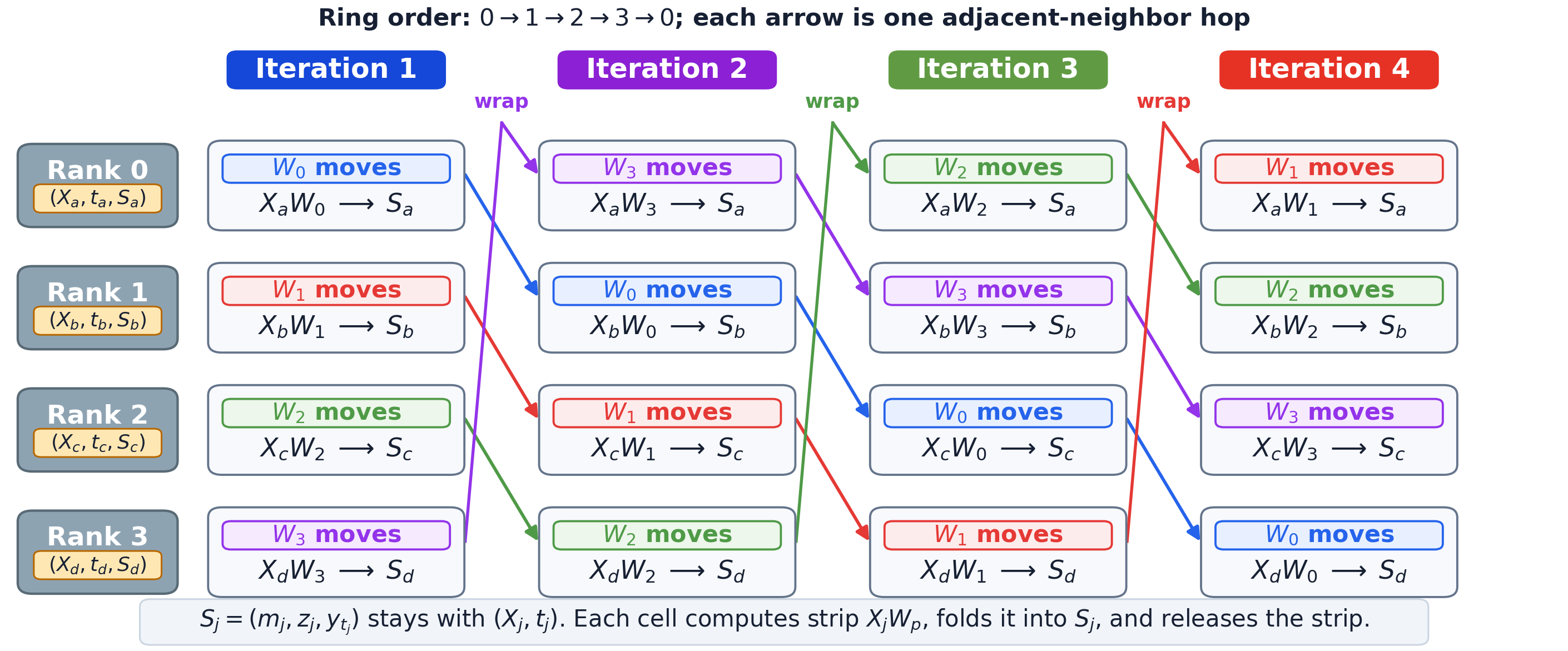}
  \caption{The move-weights schedule at $P=4$.  Rank $r$ retains its batch $(X_r,t_r)$ and
  running statistics $S_r$ while each vocabulary block $W_p$ advances one hop per
  iteration.  A meeting computes $X_rW_p$, folds the strip into $S_r$, and releases it,
  so after $P$ meetings every $S_r$ is complete on its home rank and forward needs no
  return hop.}
  \label{fig:ringtp-move-weights}
\end{figure}

The benchmark gives each rank a distinct local batch with $N=16{,}384$, uses $H=7{,}168$ and $V=200{,}000$ in FP32, and measures hidden states to target-token log-probabilities followed by backward. The standard reference stores the full projection weight and materializes full logits and log-softmax.

\begin{table}[ht]
\centering
\small
\caption{Peak allocation for the vocabulary projection as the \ringtp group grows.
Projection peaks include target-token log-probability computation and backward;
the increment subtracts the measured resident weight-plus-input footprint.  Changes are
relative to the standard dense projection.  Matched forward latency at the same shape is
$989.8$ against $1035.7$\,ms at $P=4$ and $1001.2$ against $1052.2$\,ms at $P=8$.}
\label{tab:ringtp-memory-benchmark}
\begin{tabular}{@{}c L{0.30\textwidth} R{0.19\textwidth} R{0.19\textwidth} R{0.17\textwidth}@{}}
\toprule
$P$ & Metric & Standard & \ringtp & Change \\
\midrule
\multirow{4}{*}{4}
& Projection weight & $5.3406\gib$ & $1.3351\gib$ & 75.0\% lower \\
& Resident weight + input & $5.8409\gib$ & $1.8353\gib$ & 68.6\% lower \\
& Projection F+B peak & $42.4620\gib$ & $11.5194\gib$ & 72.9\% lower \\
& Increment over resident & $36.6212\gib$ & $9.6841\gib$ & 73.6\% lower \\
\midrule
\multirow{4}{*}{8}
& Projection weight & $5.3406\gib$ & $0.6676\gib$ & 87.5\% lower \\
& Resident weight + input & $5.8409\gib$ & $1.1678\gib$ & 80.0\% lower \\
& Projection F+B peak & $42.4620\gib$ & $\mathbf{7.3222\gib}$ & \textbf{82.8\% lower} \\
& Increment over resident & $36.6212\gib$ & $6.1545\gib$ & 83.2\% lower \\
\bottomrule
\end{tabular}
\end{table}

\begin{table}[ht]
\centering
\small
\caption{Matched forward latency at the shape of
\cref{tab:ringtp-memory-benchmark}.  Local batch is per GPU.}
\label{tab:ringtp-forward-benchmark}
\begin{tabular}{@{}rrrrr@{}}
\toprule
Local batch & $P$ & Standard (ms) & \ringtp (ms) & Slowdown \\
\midrule
16,384 & 4 & 989.8 & 1035.7 & 4.6\% \\
16,384 & 8 & 1001.2 & 1052.2 & 5.1\% \\
\bottomrule
\end{tabular}
\end{table}
\section{\selectoffload}\label{app:sco}

\selectoffload walks the checkpointed layers in forward order and offloads each boundary that still fits in the host budget. Selected boundaries are copied asynchronously to capped pinned memory, and the device copy is freed. Backward visits layers in reverse. While layer $\ell$ recomputes from $h_\ell$, the next boundary $h_{\ell-1}$ is copied back on a separate stream, so at most two restores occupy the device. Stream events preserve producer and consumer order. An allocation that would exceed the cap is left on the device.

\begin{algorithm}[ht]
\caption{\selectoffload over the checkpointed layers of one session}
\label{alg:sco}
\begin{algorithmic}[1]
  \State {\bfseries Input:} checkpoint boundaries $h_1,\ldots,h_L$ in forward order, pinned
  host budget $B$
  \State $\mathcal S\leftarrow\emptyset$
  \For{$\ell=1,\ldots,L$}
    \If{$h_\ell$ fits in the unused part of $B$}
      \State copy $h_\ell$ to pinned host memory on the copy stream, then free the device
      copy
      \State $\mathcal S\leftarrow\mathcal S\cup\{\ell\}$
    \EndIf
  \EndFor
  \State \textbf{if} $L\in\mathcal S$ \textbf{then} start the restore of $h_L$ on the copy
  stream
  \For{$\ell=L,\ldots,1$}
    \State \textbf{if} $\ell\in\mathcal S$ \textbf{then} wait for the restore of $h_\ell$
    \State \textbf{if} $\ell{-}1\in\mathcal S$ \textbf{then} start the restore of
    $h_{\ell-1}$ on the copy stream
    \State recompute layer $\ell$ from $h_\ell$ and run its backward
    \State \textbf{if} $\ell\in\mathcal S$ \textbf{then} free the restored $h_\ell$
  \EndFor
  \State release the pinned storage of $\mathcal S$
\end{algorithmic}
\end{algorithm}

The host cap is a correctness condition rather than a comfort margin, because a rank that raised while the others entered a collective would deadlock. Setting it at or above $|\mathcal S|N_{\max}Hb$ makes the check pass on every rank independently of what the others hold.

The matched benchmark uses gpt-oss-20b in BF16 on one eight-H200 node with sequence and expert parallelism of degree eight. Its 24-layer decoder exposes 47 checkpoint boundaries across the hidden and residual streams. Every policy processes the same 556{,}432 tokens per measured step at a configured ceiling of 557{,}056 and runs ten complete steps, the first two excluded as warmup, with the same stateless optimizer on both sides.

\begin{table}[ht]
\centering
\small
\setlength{\tabcolsep}{2pt}
\caption{Matched \selectoffload budget sweep at 557{,}056 configured tokens.  Logical
payload is selected boundary storage and the pinned column includes host allocator size
class rounding; both are per rank.  Peak node RAM includes the common process baseline.
Memory columns are GiB, step time is seconds, throughput is tokens/s/GPU.}
\label{tab:sco-matched-benchmark}
\begin{tabular}{@{}L{0.10\textwidth} R{0.10\textwidth} R{0.13\textwidth} R{0.12\textwidth} R{0.12\textwidth} R{0.13\textwidth} R{0.10\textwidth} R{0.11\textwidth}@{}}
\toprule
Policy & Bound. & Logical & Pinned & Peak HBM & Node RAM & Step & Throughput \\
\midrule
Off & 0 / 47 & 0 & 0 & 139.790 & 402.517 & 26.3412 & 2,641 \\
$8\gib$ & 21 / 47 & 7.82--7.84 & 10.5 & 133.546 & 487.990 & 25.8415 & \textbf{2,692} \\
$16\gib$ & 42 / 47 & 15.63--15.69 & 21.0 & 125.677 & 573.473 & 25.8806 & 2,687 \\
Full & 47 / 47 & 17.49--17.55 & 23.5 & \textbf{123.728} & 593.410 & 25.8844 & 2,687 \\
\bottomrule
\end{tabular}
\end{table}

\begin{table}[ht]
\centering
\small
\caption{Capacity on the separate 32{,}768-token search grid.  ``Largest clean'' has no
allocator-cache-flush warning; ``largest completion'' finishes all ten steps with
warnings.  The adjacent first failure brackets the marginal boundary, and gain is computed
between marginal completions.}
\label{tab:sco-capacity-benchmark}
\begin{tabular}{@{}L{0.11\textwidth} R{0.15\textwidth} R{0.16\textwidth} R{0.15\textwidth} R{0.13\textwidth} R{0.15\textwidth}@{}}
\toprule
Policy & Largest clean & Largest completion & First failure & Marginal gain & Failed alloc. \\
\midrule
Off & 458,752 & 557,056 & 589,824 & -- & $27.61\gib$ \\
$8\gib$ & 557,056 & 589,824 & 622,592 & 5.88\% & $29.03\gib$ \\
$16\gib$ & 557,056 & 622,592 & 655,360 & 11.76\% & $30.59\gib$ \\
Full & \textbf{622,592} & \textbf{655,360} & 688,128 & \textbf{17.65\%} & $32.17\gib$ \\
\bottomrule
\end{tabular}
\end{table}

A session's boundaries all come from one micro-batch, so its pinned total inherits that micro-batch's length. Each session releases its pinned storage on completion, which is what makes the cap per-session rather than cumulative; otherwise the sessions comprising one update would accumulate.

\section{\streamadam}\label{app:streamadam}

A bucket co-groups a parameter shard, its fp32 master value and two moments, its final gradient, and the refreshed bf16 working weight. The fused kernel applies the same bias correction, weight decay, and update order as the reference implementation \citep{loshchilov2019decoupled}, and buckets are built once from parameter order, so ownership does not change between steps. Beginning a bucket during backward instead of after it additionally requires the bucket's gradient to be final and any cross-bucket transform to be known; global gradient clipping supplies the latter only once the global norm exists. The per-bucket and predictive clipping modes that lift this constraint change the update rule and sit outside the exactness claim.

\begin{algorithm}[ht]
\caption{\streamadam on one rank}
\label{alg:streamadam}
\begin{algorithmic}[1]
  \State {\bfseries Input:} host-resident fp32 master weights and moments, final gradients,
  bucket size $\beta$, staging slots $s$
  \State partition the rank's parameter shard into buckets $g=1,\ldots,G$ of at most
  $\beta$ parameters
  \For{$g=1,\ldots,G$}
    \State wait for a free staging slot and take it
    \State on the transfer stream, copy bucket $g$'s master weight, moments, and gradient
    into the slot
    \State on the compute stream, apply the fused AdamW update and refresh the bf16 working
    weight
    \State on the write-back stream, copy the updated fp32 master weight and moments to the
    host
    \State return the slot to the free pool when its write-back event completes
  \EndFor
\end{algorithmic}
\end{algorithm}

\subsection{Bucket and Slot Sweep}

All measurements use gpt-oss-20b on eight H200 GPUs, against the AVX CPU Adam kernel of ZeRO-Offload \citep{ren2021zerooffload} at $3.95$\,s per optimizer step. The selected 100M-parameter, two-slot configuration averages $1.93$\,s (\cref{tab:main:streamadam}). Two axes are swept: the requested bucket size, which sets how many parameters cross the link at once, and the number of staging slots, which sets how deep the transfer queue runs.

\begin{table}[ht]
\centering
\small
\setlength{\tabcolsep}{5pt}
\caption{Bucket-size and slot-count sweep.  Times are mean optimizer-step latency and
staging is resident memory per GPU.  The 100M bucket wins on both axes, and two slots stay
within $1.1\%$ of three at every bucket size while using a third less staging memory.  A
deeper queue does not help: step time is flat from two to five slots while each additional
slot adds one bucket's worth of staging.}
\label{tab:streamadam-geometry-sweep}
\begin{tabular}{@{}rrrrrr@{}}
\toprule
Requested bucket & Slots & Groups & Actual maximum & Time (s) & Staging (GiB) \\
\midrule
100M & 2 & 26 & 142.076M & \textbf{1.930} & \textbf{4.234} \\
100M & 3 & 26 & 142.076M & 1.951 & 6.351 \\
150M & 2 & 14 & 208.431M & 2.204 & 6.212 \\
150M & 3 & 14 & 208.431M & 2.225 & 9.318 \\
200M & 2 & 14 & 244.938M & 2.240 & 7.300 \\
200M & 3 & 14 & 244.938M & 2.252 & 10.950 \\
250M & 2 & 10 & 308.586M & 2.282 & 9.197 \\
250M & 3 & 10 & 308.586M & 2.282 & 13.795 \\
\bottomrule
\end{tabular}
\end{table}

\end{document}